\documentclass[12pt, a4paper]{article}

\usepackage[top=1in, bottom=1in, left=1in, right=1in]{geometry}

\usepackage[utf8]{inputenc}

\usepackage{setspace}
\usepackage{amsmath, amssymb, amsthm, amsfonts, mathtools, bm, mathrsfs}

\usepackage{graphicx}
\usepackage{float}
\usepackage{subcaption}
\usepackage{adjustbox}
\graphicspath{{figures_main/}}
\usepackage{ragged2e}

\usepackage{booktabs}
\usepackage{multirow}
\usepackage{threeparttable}
\usepackage{longtable}
\usepackage{tabularx}
\usepackage{array}
\usepackage{dcolumn}
\newcolumntype{d}[1]{D{.}{.}{#1}}

\usepackage{natbib}
\usepackage{hyperref}
\hypersetup{
    colorlinks=true,
    linkcolor=FireBrick,
    citecolor=FireBrick,
    urlcolor=blue
}
\usepackage[nameinlink,noabbrev]{cleveref}

\usepackage{appendix}
\usepackage{color}
\usepackage[dvipsnames,svgnames]{xcolor}
\usepackage{enumerate}
\usepackage{nth}
\usepackage{rotating}
\usepackage{lscape}
\usepackage{morefloats}
\usepackage{comment}
\usepackage{verbatim}
\usepackage{todonotes}
\usepackage{pdfpages}
\usepackage{optidef}
\usepackage{eucal}
\usepackage{footmisc}
\usepackage{placeins}
\usepackage{caption}
\usepackage{subcaption}
\allowdisplaybreaks[1]

\begin{document}
\pagenumbering{gobble}	

\title{\Large\textbf{\color{blue} Gas supply shocks, Uncertainty and Price setting: Evidence from Italian Firms}\thanks{\scriptsize We thank Alessandro Guerriero, Mirela Miescu and the participants to the CLIMADAP Workshop (2025, Lancaster) and Developments in Macroeconomics and Macrofinance Workshop (2025, Leipzig).}

}

\author{\normalsize
\begin{minipage}{0.45\textwidth}
    \centering
    Giuseppe Pagano Giorgianni\thanks{\scriptsize Sapienza University Rome, E-mail address: giuseppe.paganogiorgianni@uniroma1.it} \\
\end{minipage}
}

\date{\small\today}
\maketitle

\begin{center}
\href{https://drive.google.com/file/d/1oCCfLsLk5wOBVyHpkjQvUh7uSgFBi1QR/view?usp=drive_link}{Most recent version}
\end{center}
\vspace{.5cm}

\begin{abstract}

\noindent 
\footnotesize

This paper examines how natural gas supply shocks affect Italian firms’ pricing decisions and inflation uncertainty using quarterly survey data from the Bank of Italy’s Survey on Inflation and Growth Expectations (SIGE). We identify the shock using a bridge Proxy-SVAR approach that exploits high-frequency information and maps it into lower-frequency dynamics. Gas supply shocks are quickly transmitted to spot electricity prices and generate persistent effects. Additionally gas supply shocks raise firms’ prices as well as inflation uncertainty, and affect the intensive and extensive margins of price setting, with heterogeneous effects across sectors. Finally, we uncover substantial nonlinearities using state-dependent local projections: under low uncertainty recessionary forces dominate, leading firms to reduce prices below baseline.
\end{abstract}

\vspace{.5cm}
\noindent\textbf{Keywords:} Gas supply shocks, Price Setting, Inflation Uncertainty, Firms Expectations, Local Projections, Bridge Proxy-SVAR\\

\noindent\textbf{JEL Codes:E31, D840, Q41} 
	\pagebreak

\pagenumbering{arabic}

\section{Introduction} \label{sec:1}

The period that follows the Covid-19 pandemic has seen an unprecedented rise in inflation in European economies. The first part of the post pandemic recovery has been characterized by strong political tensions which led to massive disruptions in natural gas supplies to European countries, raising the attention on inflationary role of supply shocks affecting this important commodity.
Natural gas has several desirable properties for electricity generation. First, it emits less pollutants than other common fossil fuels, which increased its importance in the perspective of the achievement of the European climate objectives. Second, it can replace alternative fuels in already existing power stations, and allows for a greater flexibility in dealing with fluctuations in electricity demand.
This has led to a growing reliance on natural gas for electricity generation in several European countries, as gas-fired power plants have progressively taken on a baseload role, compensating for the decline in nuclear capacity and ensuring system reliability in the presence of intermittent renewable generation. Additionally, natural gas is a key input for many manufacturing industries, such as the chemical and pharmaceutical sectors, and overall represents about 30\% of the industry sector energy consumption in the European Union. Hence, disruptions in the supply of natural gas can affect prices through both of these channels. Despite the increased dependence on this fuel, Europe lacks large natural gas reserves and therefore relies heavily on imports. The Italian economy is a clear example of this. Despite having access to limited natural gas reserves, Italy has steadily reduced domestic production since 2000 and has remained highly dependent on imports. Natural gas accounts for about 40\% percent of electricity generation in the country. We show that shocks to the supply of natural gas affect positively both the Italian electricity price and the growth rate of firms' prices. We identify the shock by adopting the Bridge Proxy-SVAR (B-PSVAR) approach suggested by \cite{gazzani2020bridge}. The procedure consists in two steps. First, we identify the structural shock in a daily high frequency SVAR. Then, we aggregate the shock at lower frequencies and, after testing that it satisfies the exogeneity conditions suggested by \cite{FORNI2014124}, we use it as an instrument for the identification in a lower frequency model. We evaluate the effect of this shock on firms' pricing behavior and expectations using data on Italian firms from the Survey on inflation and growth expectations (SIGE), conducted by Bank of Italy. We compare our identification with the one obtained following \cite{kanzig2021macroeconomic} Proxy-VAR approach, where we construct the external instrument by exploiting news about natural gas supply disruptions \citep{alessandri2025natural, colombo2025understanding}. Additionally, we evaluate the presence of nonlinearities in the pass-trough by using state dependent local projections \citep{ramey2018government, F2021}.

\paragraph{Related literature.} 

We draw from a vast literature about macroeconomic outcomes of energy related shocks. A large number of papers study the macroeconomic effects of oil shocks \citep{hamilton1983oil, kilian2009not, caldara2019oil, conflitti2019oil, kilian2022oil}. 
Our work is closely related to the recent papers which study the effect of gas supply shocks, focusing on the recent inflationary surge in European economies. \cite{alessandri2025natural} identify a natural gas supply shock using daily news on European gas market as an instrument. An similar approach is used by \cite{colombo2025understanding}, which also identify a gas demand shocks for the US and for the Euro Area by exploiting anomalies in temperature variations.
\cite{boeck2025natural} identify a gas supply shock for the Euro Area by using a combination of sign and zero restrictions, while \cite{lopez2025pass} use sign and narrative restriction. Both of them estimate the pass through of gas supply shocks to inflation. \cite{adolfsen2024gas} use a BVAR with sign and narrative restrictions to identify shocks driving the natural gas market in the EU. They document that the pass through is heterogeneous, and depends on type of gas related shocks hitting the economy. Additionally, they document the presence of non-linearities related to labor market tightness. Similarly, \cite{guntner2024sudden} evaluate the effects of natural gas supply and demand shocks and study the effect of recent natural gas supply disruptions in the German market. \cite{casoli2024energy} study the interaction between oil/gas shocks and their effects on inflation in the Euro area. 
Most of these studies evaluate the pass through of supply distruptions on inflation by looking at the effects on HICP and its components. We instead study how firm rect to the shock in terms of their pricing strategies by using the reported annual rate of changes in their prices. Finally, \cite{granziera2025five} study the effect of natural gas supply disruptions on the first moment of firms’ inflation expectations. We extend their analysis to the second moment.

Secondly, our work relates to the literature on uncertainty \citep{bloom2007uncertainty, Bloom2009}, especially with regard to that related to measuring uncertainty in agents' expectations \citep{binder2017measuring, Jurado2015, RossiTatevik, Manski2018} and to quantifying the effects of uncertainty on the macro economy \citep{Bloom2018, ascari2022non, GGCK2024, fasani2025belief}. With respect to the previously cited papers, we focus on inflation uncertainty of firms, and therefore we are close to \cite{yotzov2023firm}, which estimates inflation uncertainty for UK firms. 

The rest of the paper is organized as follows. In \cref{sec:2} we describe the Italian gas market infrastructure and the main features in the Survey of Inflation and Growth Expectation (SIGE), which constitute our main data source of information for firm level micro data. We then describe the aggregate statistics obtained from the survey and that we use for conducting our empirical analysis, which is described in \cref{sec:3}. \Cref{sec:4} concludes.

\section{Italian Gas Infrastructure and Firm-Level Expectations Data} 
\label{sec:2}

\paragraph{Italian gas infrastructure}

Natural gas enters Italy through several entry points, the most important of which are located in Tarvisio, in the north of the country, and Mazara del Vallo, in Sicily. Tarvisio mainly receives natural gas from Russia through the Trans Austria Gasleitung (TAG) pipeline, while Mazara del Vallo receives natural gas from Algeria via the TransMed pipeline. Other important entry points include the Gries pass, which receives natural gas from Northern Europe via Switzerland through the Transitgas pipeline; Gela, which is the entry point for Libyan gas through the Greenstream pipeline; Melendugno, which receives gas from Azerbaijan via the Trans Adriatic Pipeline (TAP); and Gorizia, which receives gas from Slovenia through the Interconnector pipeline. 

\begin{figure}[H]
    \centering
    \begin{minipage}{0.9\linewidth}
    \begin{subfigure}[h]{1\textwidth}
    \includegraphics[width=\textwidth]{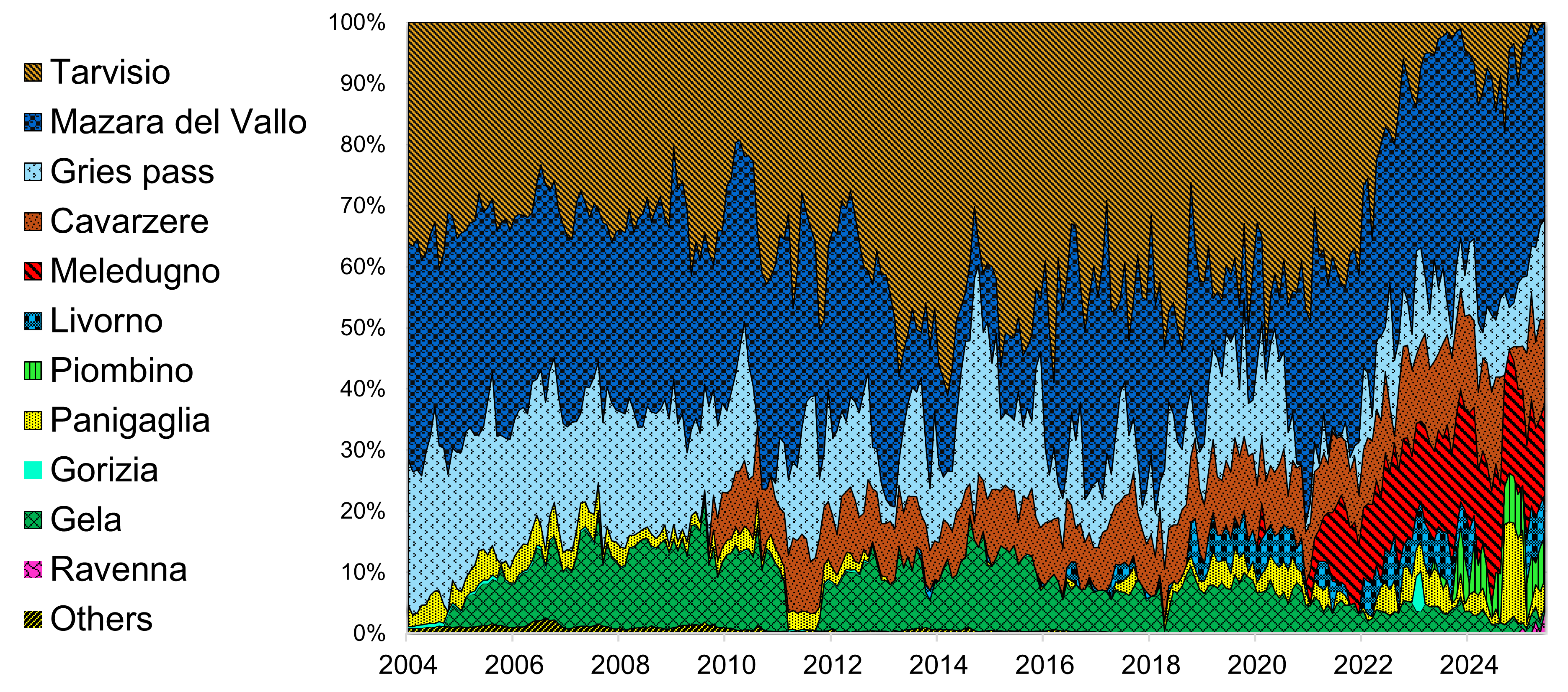}
    \end{subfigure}
       \caption{\scriptsize \justifying \textbf{Italian gas imports by entry points, as a percentage of total imported gas.} The figure reports the Italian gas imports by entry point, expressed as a percentage of total imports. Sample: 2004M1 – 2025M6. Source: Ministero dell’ambiente e della sicurezza energetica.}
    \label{figura gas_imports}
    \end{minipage}
\end{figure}

As shown in \autoref{figura gas_imports}, Tarvisio has been the most important gas hub in the country in terms of volumes, as Russia was Italy’s main supplier of natural gas. However, following the Russian invasion of Ukraine in February 2022, gas flows through this hub sharply declined as imports from Russia were progressively curtailed, leading to a significant reconfiguration of Italy’s gas supply mix. Italy responded to the reduction in inflows at Tarvisio primarily through increased imports from Azerbaijan via the Trans Adriatic Pipeline (TAP), which entered into commercial operation at the end of 2020 and became fully operational in 2021. In addition, liquefied natural gas (LNG) has played an increasingly important role in diversifying supply since 2022. Italy imports LNG through several regasification terminals, including Panigaglia and Livorno, as well as the floating storage and regasification units in Piombino and Ravenna, which became operational in 2023 and have helped offset the reduction in pipeline gas from Russia. Hence, with respect to other large European economies which relied heavily on direct pipeline imports from Russia, Italy entered the 2022 energy crisis with a more diversified gas supply structure, including LNG regasification capacity and multiple pipeline routes, which mitigated the impact of the disruption in Russian gas flows.

\paragraph{Data: Survey on Inflation and Growth Expectations}
This section illustrates the features of the Italian Survey on Inflation and Growth expectations (SIGE)\footnote{\url{https://www.bancaditalia.it/pubblicazioni/indagine-inflazione/}}, which constitutes our primary data source for firm expectations and price-setting decisions. 

As one of the longest-running firm-level expectation surveys in a G7 country, SIGE provides a rich source of information on both expectations formation and pricing behavior. The survey is conducted by the Bank of Italy at a quarterly frequency, starting from the end of 1999. It delivers a rotating panel, collecting different kinds of information from Italian firms with at least 50 employees and belonging to the manufacturing and construction sector.\footnote{Since 2013, also firms belonging to the construction sector have been included.} 
The interviews are conducted during the last month of each quarter. Among others, firms are asked to provide a point estimate for their year on year inflation expectations, and the expected change in their own prices over the course of the year. Furthermore, firms report the average change in their own realized price over the last year, allowing a comparison between expectations and outcomes.\footnote{Additionally, firms respond to a range of categorical question related to what are the main factors that will affect their own prices over the course of the next year. In this case, responses are on a scale giving information about both direction (downward or upward pressure) and intensity (ranging from strong to modest).} Apart from that, we exploit this rich data source to get aggregate measures of:  (i) firms' intensive and extensive margins of price adjustment, where the intensive margin provide a proxy of the intensity of the price changes and the extensive margin of the portion of the sample which updates in each quarter; (ii) a measure of firms' uncertainty about future inflation. 

It is important to note that firms are informed about the previous month level of inflation in the moment they are interviewed, as this information is reported in the survey questionnaire. The question about firms' expected level of prices remained mostly unchanged since the start of the survey, with the following rough formulation: 

\begin{quote}
\textit{Last month, the consumer inflation rate, measured by the twelve-month change in the Harmonised Index of Consumer Prices, was equal to XX\% in Italy and to YY\% in the euro area. What will be the consumer inflation rate in Italy in 12 months?}
\end{quote}

As shown by \cite{CGR2020}, this affects the way in which the firms' answer about their inflation expectations, but the effects on firms' pricing decisions should be negligible and not statistically significant.\footnote{Since 2012Q3, the sample size has been increased and firms have been randomly assigned to two groups: one receiving updated information about current inflation, and the other receiving no such information. Due to the length of the sample size, the aggregate statistics we report are those related to firms which have been updated about the current level of prices.}

\begin{figure}[H]
  \centering

  \begin{subfigure}[t]{0.49\textwidth}
      \caption{Price changes}
    \includegraphics[width=\textwidth]{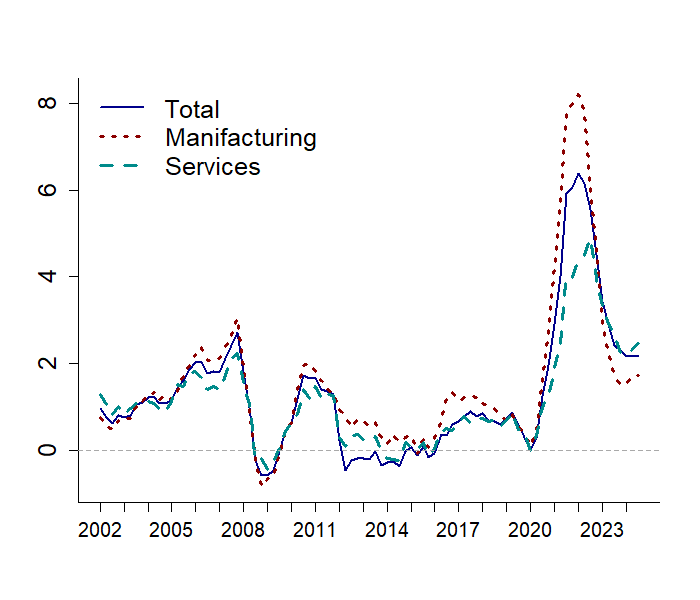}
  \end{subfigure}
  \hfill
  \begin{subfigure}[t]{0.49\textwidth}
    \caption{Expected price changes}
    \includegraphics[width=\textwidth]{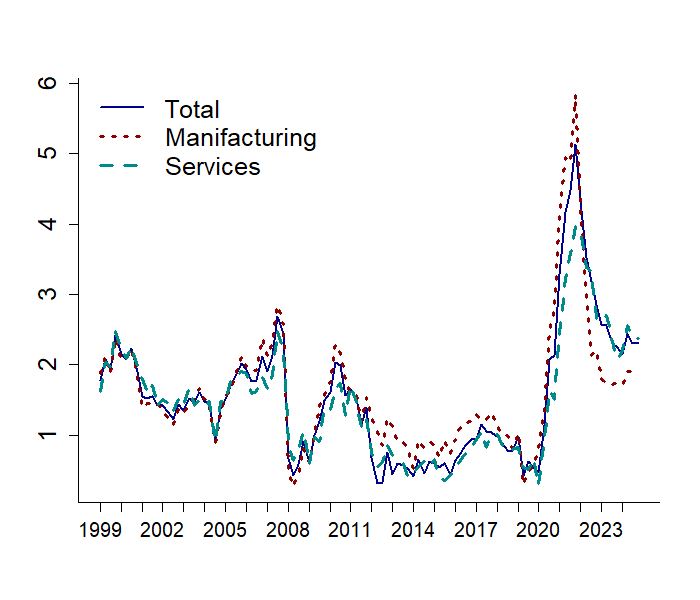}
  \end{subfigure}

  \caption{\scriptsize \justifying 
  \textbf{Price realizations and expectations across sectors.}  
  The figure plots average price changes over the last 12 months (left) and the average expected price changes over the next 12 months (right) for Italian firms in the manufacturing (red dotted line) and services sectors (lightblue dashed line), compared with the overall series (blue solid line).  
  Source: Bank of Italy, Survey on Inflation and Growth Expectations.  
  Sample: 1999:Q4–2025:Q2.}
  \label{figura 9b}
\end{figure}

Realized and expected price changes, reported in \autoref{figura 9b}, display very similar dynamics. Prior to the Great Recession, average price changes ranged between 1 and 2.5\%. In 2008, they turned sharply negative and, after a brief rebound, they remained close to zero throughout the zero lower bound (ZLB) period. After 2020, both expected and realized price changes rose sharply, reaching historical highs. 
Marked differences emerge across sectors. Price changes in manufacturing display higher volatility and larger amplitudes than in services, while price changes in services evolve more smoothly, particularly in the post-pandemic period. After 2022, price changes in manufacturing decline more abruptly, whereas price changes in services adjust more gradually and remain elevated for a longer period. A similar assessment is also valid in the case of expected price changes. 

We exploit firms’ answers on how much their prices changed over the previous 12 months to construct proxies of the extensive and intensive margins of price adjustment. The extensive margin is defined as the share of firms reporting a non–negligible year-on-year change in prices, which provides a survey–based measure of the incidence of price adjustment. The intensive margin is the average reported year-on-year price change among firms that adjusted. These indicators do not track the precise timing or the number of price changes within the year, but they summarize whether firms adjusted at least once over the past 12 months and by how much, and therefore offer meaningful proxies for the underlying price-setting margins. 

Let $\Delta p_{i,t}^{yoy}$ denote the year-on-year percentage change in the price reported by firm $i$ in quarter $t$. We define our proxy of the extensive margin $F_t$ as the share of firms reporting a non-negligible price change: 

\begin{equation*}
    F_t = \frac{1}{N_t}\sum_{i=1}^{N_t} \mathbf{1}\!\left(|\Delta p_{i,t}^{yoy}| \ge \varepsilon\right)
\end{equation*}

where $\varepsilon$ is a small threshold used to filter out reporting noise. The intensive margin $I_t$ is instead defined as the average reported year-on-year price change among adjusting firms:

\begin{equation*}
    I_t = \mathbb{E}\!\left[\Delta p_{i,t}^{yoy} \,\middle|\, |\Delta p_{i,t}^{yoy}| \ge \varepsilon\right].
\end{equation*}

\begin{figure}[H]
  \centering

  \begin{subfigure}[h]{0.49\textwidth}
    \caption{Extensive margin}
    \includegraphics[width=\textwidth]{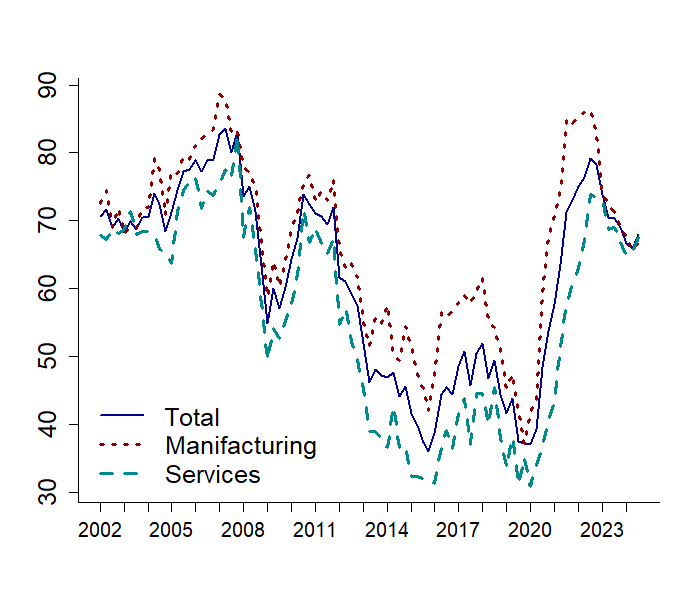}
  \end{subfigure}
  \hfill
  \begin{subfigure}[h]{0.49\textwidth}
    \caption{Intensive margin}
    \includegraphics[width=\textwidth]{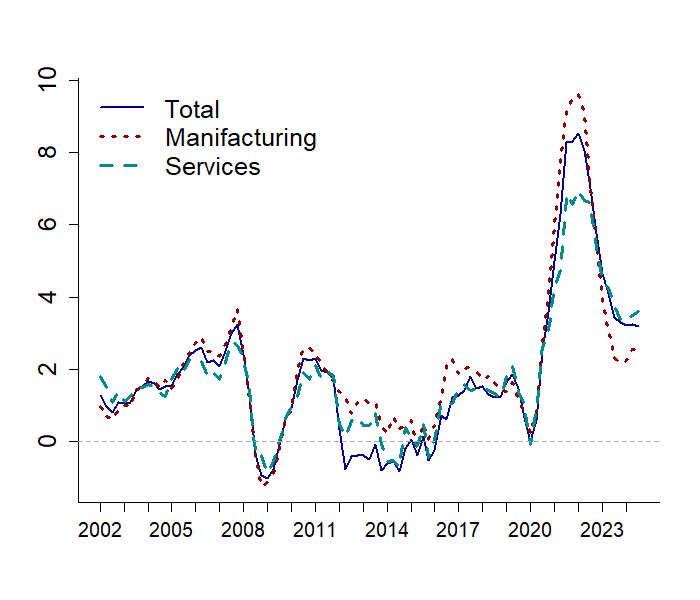}
  \end{subfigure}

  \caption{\scriptsize \justifying \textbf{SIGE intensive and extensive margins} The figure plots the proxies for the extensive (left) and extensive (right) margins from the SIGE survey responses about the reported price changes in the last 12 months for Italian firms. Source: Bank of Italy, Survey on inflation and growth expectations. Sample: 1999:Q4–2025:Q2.}
  \label{figura 13}
\end{figure}

\noindent Under this definition, the unconditional mean of firms' year-on-year price changes satisfies: 

\begin{equation*}
\mathbb{E}[\Delta p^{yoy}_{i,t}] = F_t\, I_t,
\end{equation*}
The two series, represented by the solid blue lines in \autoref{figura 13}, exhibit cyclical patterns that are consistent with the micro-pricing evidence and provide a characterization of the price adjustment behavior of Italian firms over time. 
For both of them, we observe positive comovements with realized inflation. Before 2008, roughly 80\% of firms in the sample adjusted their prices at least once per year, and the magnitude of these adjustments remained moderate. In 2008, both margins collapse: fewer firms change their prices, and those that do implement substantial price cuts, marking a sharp and broad-based disinflationary episode aligned with the onset of the global financial crisis.
Following this break, the period from 2011 to 2019 is characterized by persistently muted pricing behavior. The extensive margin settles at historically low levels, while the intensive margin hovers around zero, indicating that price adjustments—when they occur—are infrequent, small, and lack a clear direction. This pattern reflects a prolonged environment of weak inflation and weak demand.
After 2020, the dynamics shift abruptly. The extensive margin rises markedly, signaling that a much larger share of firms resumes active price adjustment, while the intensive margin reaches unprecedented highs, pointing to large and predominantly upward price changes. These developments reveal a clear regime shift towards strong and widespread price increases driven by the post-pandemic recovery, the surge in energy costs, and the European gas supply shock.

Computing the price-setting margins by differentiating across sectors reveals clear patterns. First, compared with the services sector, a slightly higher share of manufacturing firms adjust their prices over the sample. The gap between the two groups widens from 2013 until the onset of the pandemic. Manufacturing firms also start increasing their prices more frequently after 2020. In 2022, almost 90\% of manufacturing firms adjusted their prices. By contrast, in the services sector the increase in the intensive margin is more gradual and contained, although still substantial.

The intensive margin is similar for both sectors for most of the sample, but after 2020 clear differences emerge: manufacturing firms adjust their prices more aggressively until 2023, after which their intensive margin decreases sharply. In contrast, price adjustments by firms in the services sector remain more contained overall, but the size of their adjustments surpasses that of manufacturing firms after 2023.

\paragraph{Firms' inflation uncertainty} 

To measure firms' inflation uncertainty, we adopt two alternative approaches. First, we proceed as proposed by \cite{binder2017measuring}, which exploits the well-documented tendency of survey respondents to provide round-number forecasts when they face greater uncertainty. This method yields an index that captures the proportion of likely uncertain respondents in each survey wave. Differently from her case, our respondents can provide decimal answers. Therefore, we need to adapt her framework to our setting by adjusting the likelihood function accordingly.
We assume that there are two type of respondents. $Type-h$ respondents are characterized by an high degree of uncertainty in their subjective probability distribution, while $type-l$ respondents by a low degree of uncertainty. $Type-h$ respondents are assumed to give point estimates for their expected level of prices coming from the set $S_h=\{0,5,10,15,20,25\}$, while low uncertainty firms are allowed to answer with decimal numbers on the support $\mathbb{R}^+$. If an answer $R_{it}\notin S_h$ we can surely classify it as $type-l$. Conversely, if $R_{it}\in S_h$ we need to estimate the probability that respondent $i$ is uncertain. We assume that the cross sectional density is the mixture of two distributions. For $type-h$ respondents, coming from $S_h$ the probability mass function is obtained by integrating the latent density over rounding bins: $\phi_h(s) = P(R_{i,t} =s|type-h) = \int_{s-\delta/2}^{s+\delta/2}p_h(x)dx$, with $j\in S_h$ and $\delta$ being the bin width.\footnote{We set $\delta=0.1$.} For $type-l$ respondents the probability density is simply $f_l(x)=p_l(y)$ with $y\in \mathbb{R}^+$ and $\phi_l(s) = \int_{s-\delta/2}^{s+\delta/2}p_l(x)dx$ for rounded observations. The cross sectional log likelihood will then be given by:

\begin{equation}
l(\theta_t) = \sum_{s=S_h} n_h(s) log[(\lambda_t \phi_h(s) + (1-\lambda_t)\phi_l(s)] + \sum_{i=1}^{N_c} log[(1-\lambda_t)f_l(y_{c,i};\mu_l,\sigma_l)]
\end{equation}

\noindent where $f_l$ is assumed to be normal, $n_h$ is the number of firms which responses belong to $S_h$, $N_c$ is the number of non rounded responses and $\theta=(\phi_h,\lambda_t,\mu_l,\sigma_l)$ is a vector of parameters .
The mixture weight $\lambda_t$ is the portion of uncertain respondents.\footnote{\cite{binder2017measuring} includes non respondents and measures the proportion of uncertain households as $U_t=\frac{\lambda_tN_t + \gamma N_t^{DK}}{N_t + \gamma N_t^{DK}}$ where $N_t$ is the total number of respondents, $N_t^{DK}$ is the number of respondents which answer is "don't know" and $\gamma$ is a parameter which measures their weight. As in the SIGE survey the question about expected inflation is mandatory, we do not have $DK-type$ respondents in our sample and hence we measure the proportion of uncertain firms by simply taking $\lambda_t$.} The vector of parameters $\theta_t$ is estimated via maximum likelihood.

Alternatively, we build an uncertainty index constructed following the methodology of \cite{Jurado2015} based on the volatility of the unpredictable component of macroeconomic dynamics, by exploiting the high-dimensional dataset of macroeconomic time series for the Italian economy provided by \cite{barigozzi2024large}.\footnote{No outlier correction is performed, and missing data are imputed using the algorithm of \cite{mccracken2016fred}.} The correlation between the two uncertainty measures is 70.1\%. 
Notably, since firms are informed about the most recent level of inflation at the time of the interview, our index isolates doubts about future inflation, abstracting from any confusion about the current state of prices.\footnote{A concern may be that building the uncertainty index only on firms who receive the information about the current level of prices may lead to an under measurement in uncertainty. In \cref{App. Uncertainty} of the Appendix, we show that the share of uncertain firms who do not receive information about current inflation when interviewed is similar both in size and temporal patterns to our baseline.} 

\begin{table}[H]
    \centering
    \begin{tabular}{lcc}
                             & \textbf{Binder (2017)}   & \textbf{Jurado et al. (2015)} \\
    \toprule
    \textbf{Standard deviation SIGE}     &  0.86               &   0.84             \\
    \textbf{Inflation volatility}        &  0.75               &   0.79             \\
    \textbf{Annual inflation}            &  0.54               &   0.53             \\
    \textbf{Economic policy uncertainty} &  0.15               &   0.12             \\
    
    \end{tabular}
    \caption{\scriptsize \justifying \textbf{SIGE Inflation Uncertainty correlation with related series.} The table reports the correlations of the inflation uncertainty index for firms, alternatiely computed by adopting the framework of \cite{binder2017measuring} and \cite{Jurado2015}, with the standard deviation of firms' inflation expectations from SIGE, the annualized HICP inflation for Italy, the volatility of realized inflation, estimated by adopting an ARIMA-GARCH model and the Economic Policy uncertainty index of \cite{baker2016measuring}. Source: Bank of Italy, Survey on inflation and growth expectations, EUROSTAT, \cite{baker2016measuring}, \cite{barigozzi2024large}. Sample: 1999:Q4–2025:Q2 \citep{binder2017measuring}; 2002:Q4–2025:Q2 \citep{Jurado2015}.}
    \label{tab:placeholder}
\end{table}

In \autoref{tab:placeholder}, we report the correlations between firms’ inflation uncertainty—estimated using both methodologies —and a set of related macroeconomic indicators. We consider the standard deviation of firms' inflation expectations, as an alternative proxy commonly used in the literature to measure uncertainty in the expectations, the ex-post annual HICP inflation volatility, estimated by adopting an ARIMA-GARCH model, annual inflation, and the \cite{baker2016measuring} economic policy uncertainty index. Both uncertainty measures are strongly correlated with forecast dispersion and ex post inflation volatility. By contrast, correlations with the level of inflation and with the Economic Policy Uncertainty index are markedly lower, suggesting that the index is not driven by inflation levels or by general policy uncertainty. Instead, it primarily reflects uncertainty about inflation dynamics, rising during periods of increased inflation volatility.

\begin{figure}[H]
  
    \centering
    \begin{minipage}{1\linewidth}
    \begin{subfigure}[h]{1\textwidth}
    \includegraphics[width=\textwidth]{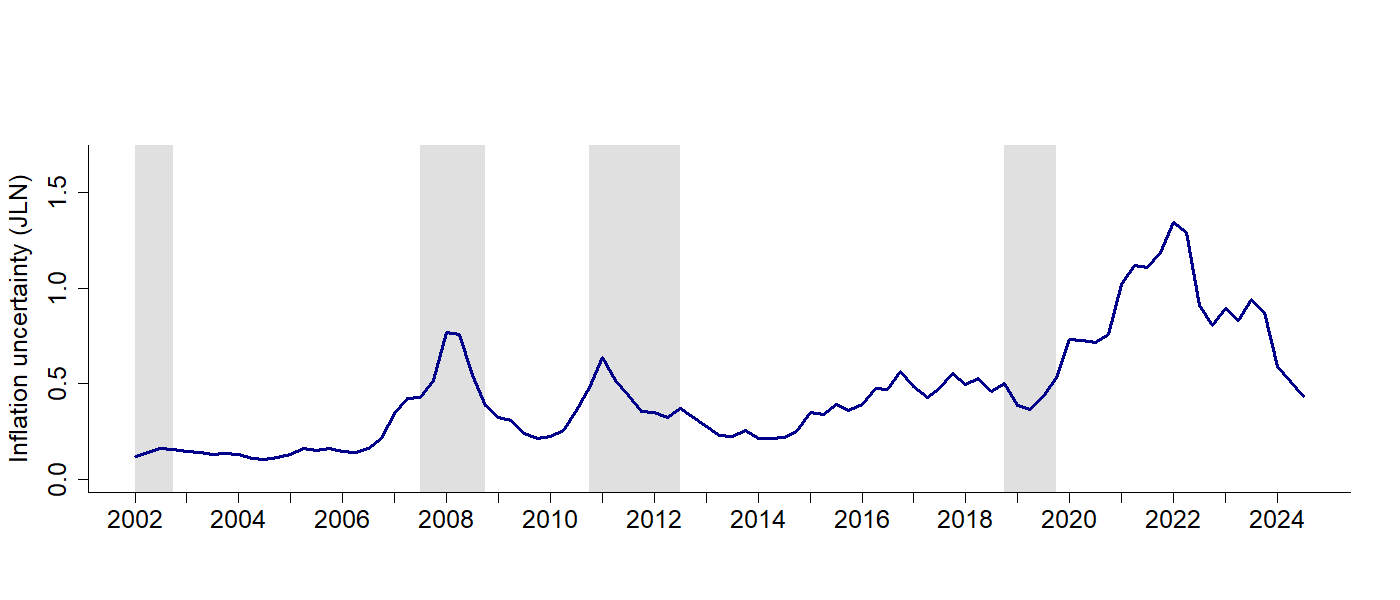}
    \end{subfigure}
       \caption{\scriptsize \justifying \textbf{Inflation Uncertainty Index.} The figure reports the inflation uncertainty index for firms computed by adopting the framework of \cite{Jurado2015}. Shaded areas represent OECD recessions. Source: Bank of Italy, Survey on inflation and growth expectations, OECD. Sample: 2002:Q4–2025:Q2.}
    \label{figura 2}
   \end{minipage}
\end{figure}

\autoref{figura 2} reports the inflation uncertainty based on the \cite{Jurado2015} approach. During the first part of the sample, uncertainty remains low, reflecting stable price dynamics. It increases during the Global Financial Crisis and the European sovereign debt crisis. Uncertainty rises again in the last quarter of the turbulent 2020 and increases further in the first quarter of 2022, following Russia’s invasion of Ukraine and the onset of the European energy crisis. It reaches its highest level in the fourth quarter of 2022, when Italian inflation attains its peak since the adoption of the Euro, underscoring widespread concerns about the persistence and diffusion of price pressures.

\section{Empirical Analysis} \label{sec:3}

The aggregate series about firms' pricing and inflation expectations we recover from the SIGE microdata happen to be at the quarterly frequency. This poses a challenge for the strategies commonly adopted in the literature for the the identification of natural gas supply shocks. To overcome this limitation, we rely on the bridge Proxy-SVAR approach suggested by \cite{gazzani2020bridge}. The procedure consists in three steps. (i) We identify the shock using a high-frequency VAR estimated at the daily frequency. Working at high frequency mitigates the distortions induced by temporal aggregation and allows identification to rely on weaker timing assumptions than those typically imposed in low-frequency systems, where contemporaneous restrictions span an entire month or quarter. In our case, identification in the daily system follows the methodology proposed by \cite{Kanzig2020}, which exploits high-frequency movements to isolate unexpected innovations. (ii) After extracting the daily shock and aggregating it to the monthly or quarterly frequency, we test its invertibility, following \cite{FORNI2014124}. This assesses weather the high frequency VAR is specified in an informationally sufficient manner. (iii) The aggregated shock is then used as an external instrument in a lower frequency VAR, containing macroeconomic endogenous variables of interest. 
This approach offers several advantages. First, it allows us to identify natural gas shocks at a daily frequency, where financial markets react immediately to news and contemporaneous feedback from real economic variables is limited, thereby requiring weaker identifying assumptions. Second, the bridge framework provides a coherent mapping of these high-frequency shocks into lower-frequency macroeconomic dynamics, ensuring consistency across frequencies. Third, it enables the analysis of quarterly firm-level outcomes, such as pricing behavior and inflation expectations, while preserving the identification achieved at higher frequency.  

\subsection{Identification strategy}\label{subsec:identification}

\subsubsection{High frequency instrument construction}

We build the daily instrument following the procedure suggested by \citeauthor{kanzig2023unequal} (\citeyear{kanzig2021macroeconomic}, \citeyear{kanzig2023unequal}). Our strategy to identify the narrative news is analogous to the one adopted by \cite{alessandri2025natural} and \cite{colombo2025understanding}. 
We take from Reuters, collected through LSGE refinitiv, about interruptions of gas supplies to Europe which are highly likely to be exogenous to the economy and unexpected by the markets. We search for keywords such as "outages", "interruptions" and "unplanned maintainance". Hence, the daily events we consider may include episodes of unanticipated interruptions caused by terrorist attacks, failures and outages to natural gas production facilities or pipelines, such as the explosion that hit the Baumgarten natural gas facility in Austria in 2017 causing the interruption of the flow of gas to the Italian natural gas entry point in Tarvisio, and the fire in the Freeport liquefied natural gas facility in US which reduced the supply of LNG to Europe. 
Additionally, for the 2022 period we include some announcements from Russia which have been perceived as a surprise from the market and which contain news about reduction in the supply of natural gas.\footnote{As an example, the announcement by Putin that Russia was going to stop the deliveries of natural gas, unless the payments were made in robles in March 2022 concretely reduced the amount of purchasable commodity for countries which could not or did not want to comply with the request, and has been followed by a positive movement of the Dutch TTF \citep{alessandri2025natural}}

\begin{figure}[H]
  
    \centering
    \begin{minipage}{0.9\linewidth}
    \begin{subfigure}[h]{1\textwidth}
    \includegraphics[width=\textwidth]{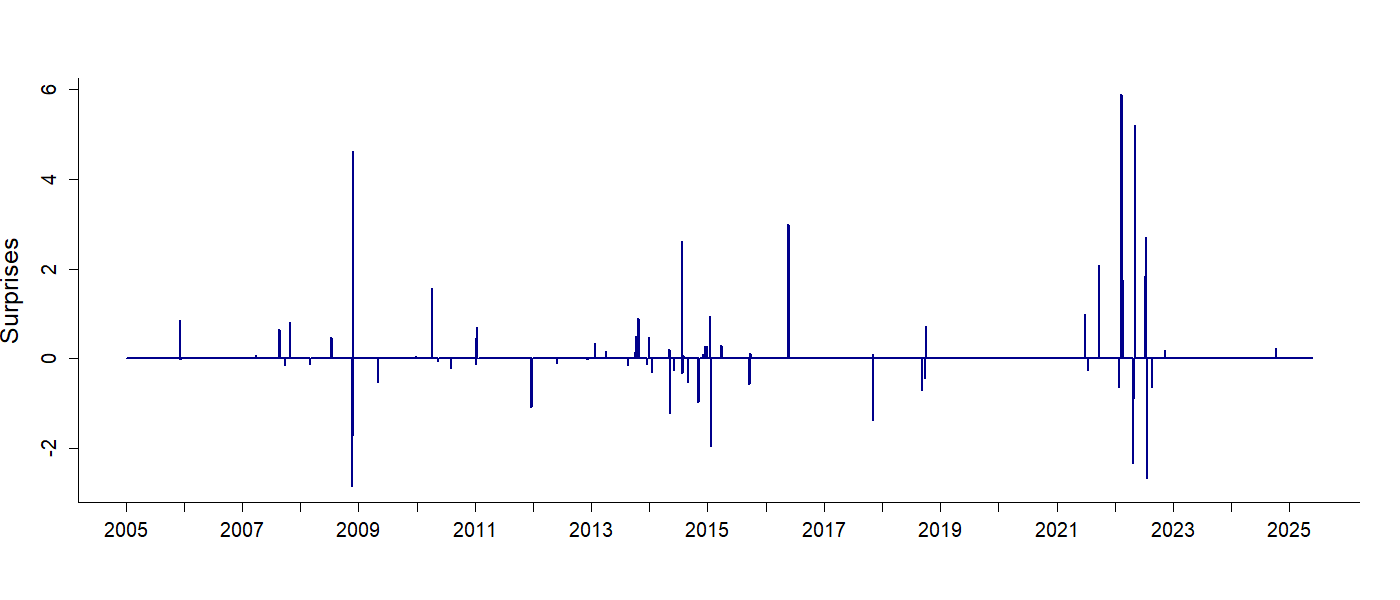}
    \end{subfigure}
       \caption{\scriptsize \justifying \textbf{News based narrative instrument.} The figure reports the daily news based narrative instrument. Sample: 2005M1 – 2025M6.}
    \label{figura surprises}
    \end{minipage}
\end{figure}

\noindent We end up isolating 75 surprises which are reported in \autoref{tab:surprises}. We construct the daily instrument by first computing log-differences of Dutch TTF futures prices at maturities going from 1 to 12 months. We then extract their first principal component, which captures the common factor in TTF futures movements. Following the argument in \cite{Kanzig2020}, assuming the risk premium is approximately constant within the narrow event window, this common futures-price movement can be interpreted as a news-induced revision in expected natural-gas prices. 
Finally, we build the narrative instrument by taking the variation of this measure around the daily event window. 

\begin{equation*}
    \textit{Surprise}_{t,d} = F_{t,d} - F_{t,d-1}
\end{equation*}

\begin{table}[H]

\centering
\tiny
\begin{tabular}{ll}
  \toprule
   \textbf{Event date} & \textbf{Event description} \\ 
   \midrule
   18 January 2006   & Gazprom is going to cut gas supply to Serbia by 25\% \\
   19 January 2006   & A gas leak and a storm force a reduction in Norwegian oil/gas output \\
   7 May 2006        & Explosion in a gas pipeline in Ukraine \\ 
   2 October 2007    & Russia announces it may cut gas supply to Ukraine over debt \\
   8 November 2007   & Interruption in Norwegian oil/gas output due to stormy weather\\ 
   7 December 2007   & Ukraine says blast hit gas pipe\\
   9 April 2008      & Blast hits Russia gas line\\
   22 August 2008    & UK N.Sea Britannia gas field output cut\\
   1 January 2009    & Russia cuts off gas supply to Ukraine\\
   5 January 2009    & Deliveries of Russian gas to Croatia reduced by 18\% \\
   6 January 2009    & Yushchenko says that Gazprom has reduced gas supply to Europe through Ukraine on the morning of 6th January\\
   7 January 2009    & Gas supplies to Italy via TAG pipeline are substantially interrupted \\ 
   6 November 2009   & Gazprom threatens to cut Ukraine gas supply \\
   4 February 2010   & Norway gas export capacity cut slightly \\
   18 May 2010       & Norway gas output reduced due to technical problem\\
   21 June 2010      & Russia President orders Belarus gas supplies cut \\
   14 September 2010 & Blast in gas pipeline in Bulgaria \\
   4 February 2011   & Power supply problem reduces output of Kollsnes gas processing plant in Norway\\
   15 February 2011  & Gas export capacity of Kaarstoe gas plant in norway reduced due to technical problems \\
   17 February 2011  & Unexpected cut in the gas flows from the South Hook LNG plant in UK\\
   22 February 2011  & Due to protests and civil war, the Greenstream pipeline stopped delivering gas from Libya to Italy\\ 
   12 July 2012      & UK Centrica's South Morecambe gas flow cut by technical issue \\
   17 January 2013   & Supplies from Algeria to Italy fall because of a terrorist attack in Almenas Gas Field\\
   4 March 2013      & Clashes with militias in Mellitha complex in Libya completely stops the gas flows to Italy. \\
   14 May 2013       & Norway's Kaarstoe gas plant output cut by 40 mcm on Tuesday\\
   30 September 2013 & Protests in Libya cause the shut down of a gas pumping station, halting the gas flows to Italy \\
   11 November 2013  & Libyan Berbers shut gas pipeline to Italy \\
   15 November 2013  & Norway's Oseberg field outage to cut gas output for 2-3 days\\
   28 November 2013  & Outage at Shell's Ormen Lange to cut Norway's gas exports by 35 mcm/day\\
   3 December 2013   & Nyhamna's outage extended, Norway's gas output cut by 25 mcm/day\\
   23 January 2014   & Gas flows from Algeria to Italy stopped due to quality issues. Received gas had to be sent back \\
   6 February 2014   & Norway gas exports cut by 27 mcm/day due outage at Troll\\ 
   25 February 2014  & UK Centrica's North Morecambe supply cut due to technical problems\\ 
   16 June 2014      & Kiev says Russia has cut all off gas to Ukraine\\
   17 June 2014      & Blast hits gas pipeline in central Ukraine\\ 
   15 July 2014      & Norway gas output to be cut by 15 mcm/day until July 16\\ 
   1 September 2014  & Gas prices jump on Norway pipeline outage, Russia gas cut fears\\ 
   6 September 2014  & Gas output at Norway's Kollsnes, Kvitebjoern cut for 24 hours\\
   12 September 2014 & Gazprom plans to further cut gas exports to Romania\\ 
   10 October 2014   & Gas output at Norway's Kollsnes gas plant and Kvitebjoern gas field to be cut for 24 hours\\
   10 November 2014  & Norway's gas output cut by 9.3 mcm/day due to Gjoea field outage\\
   15 December 2014  & Norway's gas output further cut on Monday due to Kvitebjoern outage\\
   14 January 2015   & Troll outage to cut Norway's gas production by 117 mcm \\
   27 January 2015   & Norway's Nyhamna gas plant capacity cut by 10 mcm/day\\
   4 February 2015   & Kollsnes outage to cut Norway's gas output by 19 mcm\\
   24 February 2015  & Norway's gas flows cut further by entry SEGAL outage, Moscow threatens to cut off Kiev as gas dispute resurface \\ 
   5 March 2015      & Outages to cut Norway's gas output by 14 mcm\\
   29 April 2015     & Libya shuts down a gas field in the east due to protests \\
   8 May 2015        & Libyan gas export cut to Italy\\
   30 October 2015   & Unplanned outage at Norway Aasgard gas field\\
   6 November 2015   & Kvitebjoern outage to cut Norway's gas output by 7 mcm \\
   1 July 2016       & Gas flows to Poland from Gazprom back to normal after disruption\\
   12 December 2017  & A blast in Austrian hub of Baumgarten interrupts gas transit to Tarvisio hub in Italy\\
   13 December 2017  & Gas supply from Austrian gas hub back to normal after deadly blast\\
   18 October 2018   & Norway's Nyhamna gas plant to cut output by 7 mcm due water leak\\
   8 November 2018   & Norway's Gassco to minimise gas exports disruption to Europe after Kollsnes gas processing plant shut down\\
   13 November 2018  & Gas exports from Norway's Kollsnes cut by 11.5 mcm per day \\
   28 July 2019      & Blast hits Russian gas pipeline\\
   4 August 2021     & Algeria gas flows to Italy cut by around 25\%\\
   2 November 2021   & Ukraine says Russia's Gazprom has cut daily gas transit volume to 60 mcm\\
   8 March 2022      & Russia warns it could cut gas supplies via Nord Stream 1\\
   23 March 2022     & Putin asks European countries to pay gas deliveries in Rubles otherwise gas flows will stop \\
   31 March 2022     & Putin signs the decree asking payment in Rubles to european countries\\ 
   24 April 2022     & Russia cuts gas supply to Poland\\
   1  June 2022      & Gazprom to cut gas supplies to Denmark's Orsted\\
   8 June 2022       & Fire in the Freeport LNG facility in the US \\
   14 June 2022      & NS1 operate at limited capacity due to turbines stuck in Canada\\
   19 August 2022    & Gazprom announces that Nord Stream 1 three days shut down for maintenance on the remaining operational turbine \\
   25 August 2022    & Russia says that turbines are not repaired in Canada\\
   31 August 2022    & Flows of gas from NS1 shut down permanently \\
   3 October 2022    & Eastward gas flows via Yamal-Europe pipeline stop \\ 
   20 December 2022  & Blast hits Russia-Ukraine gas export pipeline\\
   14 January 2023   & Lithuanian pipeline blast\\ 
   15 November 2024  & Austria says Russia to cut off gas from Saturday\\

  \bottomrule
\end{tabular}

\caption{\scriptsize \justifying \textbf{Natural gas related news.} The table reports the natural gas surprises we use to construct the instrument. Source: Reuters. }
\label{tab:surprises}
\end{table}

Where $F$ indicates the principal component of the log-difference of the futures, $d$ is the day in which the news event occurs, and $d-1$ is the last trading day before the occurrence of the event.\footnote{When a news event occurs on a non-trading day, it is assigned to the first subsequent trading day.} The daily surprise series is reported in \autoref{figura surprises}. 

\subsubsection{Bridge PSVAR identification}\label{sec:BPSVAR} 

\paragraph{Daily VAR-X} We estimate a daily VAR-X which includes a set of financial, commodity, and macro-financial variables commonly used in the literature to control for global demand conditions, financial market developments, and risk sentiment. Specifically, it comprises the Dutch TTF natural gas day ahead future price, which is the most common measure for the spot natural gas price in the European market, the EURO STOXX 50 equity index, Brent and WTI crude oil prices, the EUR-USD exchange rate, the VSTOXX volatility index, the composite indicator of systemic stress (CISS) for the Euro Area and the ECB overnight deposit rate.
We complement this set of endogenous variables with electricity demand from the main gas-intensive economies in the European Union, namely Austria, Belgium, Germany, Spain, France, Italy, and the Netherlands. Among the exogenous variables instead, we consider the average temperature and irradiance in those countries.\footnote{Electricity demand, irradiance and average temperature are provided at daily frequency by the Copernicus program at \url{https://cds.climate.copernicus.eu/datasets/}. The series are seasonally adjusted using regression-based methods that remove day-of-week effects and smooth intra-annual seasonality via spline functions of the day of the year; for electricity demand, the specification allows for interactions between weekly and annual seasonal patterns.}. We set the number of lags $p$ to 90 in order to cover roughly the last 3 months. The VAR is estimated in log levels from 1 January 2004 until 30 July 2025. To identify a shock to the supply of natural gas, we adopt an high frequency identification approach \cite{kanzig2021macroeconomic}. The reduced form model is characterized by the following structure:

\begin{equation*}
    y_t = \sum_{p=1}^P A_p y_{t-p} + Bx_t + u_t
\end{equation*}

\noindent where $y_t$ denotes the vector of the $n$ endogenous variables, $x_t$ denotes the vector of exogenous variables, $A_p$, for $p = 1, \dots, P$, are the reduced-form $n \times n$ coefficient matrices, and $u_t$ denotes the vector of reduced-form residuals.
We assume that the relation between the reduced form and the structural form residuals is the following: 

\begin{equation*}
    \epsilon_t = B_0^{-1} u_t
\end{equation*}

\noindent where $B_0^{-1}$ denotes the $n \times n$ structural impact multiplier matrix. We can write the variance covariance matrix of the structural form residuals as $\Sigma = B_0^{-1} \Omega {B_0^{-1}}' $ where $\Omega = Var(\epsilon_t)$ is assumed to be the identity matrix $I$. 

We aim to identify the first structural shock $\epsilon_{1,t}$, whose contemporaneous impact on the system is given by the first column of the structural impact matrix $B_0$.
We assume that our daily narrative surprise series $z_t$ satisfies the properties of independence and relevance. Namely, we assume that: 

\begin{equation*}
    \mathbf{E}(\epsilon_{1,t}z_t) \neq 0 
\end{equation*}

\begin{equation*}
    \mathbf{E}(\epsilon_{2:n,t}z_t) = 0 
\end{equation*}

\noindent If these conditions hold, the structural impact vector associated with the shock $\epsilon_{1,t}$ can be identified up to sign and scale. It is given by

\[
s_1 = \begin{pmatrix}
s_{1,1} \\
\tilde{s}_{2:n,1}
\end{pmatrix},
\qquad
\tilde{s}_{2:n,1} = \frac{\mathbf{E}(u_{2:n,t} z_t)}{\mathbf{E}(\epsilon_{1,t} z_t)} .
\]

The scale parameter $s_{1,1}$ is set by normalization. We normalize the shock so as to correspond to a 1\% increase in the TTF day-ahead price.

To assess instrument relevance, we regress the reduced-form VAR residual of the TTF price on the instrument. The F-statistic from this first-stage regression provides a measure of instrument strength \citep{olea2021inference}, with values above 10 conventionally indicating a strong instrument.

The F-statistic from this first-stage regression provides a measure of instrument strength \citep{olea2021inference}, with values above 10 conventionally indicating a strong instrument. In our case, the F-statistic equals 15.27. We also conduct a heteroskedasticity-robust F-test, which yields a value of 66.46, further confirming that the relevance condition is satisfied. The identified daily structural shock is then given by

\[
\epsilon_{1,t} = \frac{s_1' \Sigma_u u_t}{\sqrt{s_1' \Sigma_u s_1}} .
\]

\begin{figure}[H]
    \centering
    \begin{minipage}{1\linewidth}
    \begin{subfigure}[h]{0.55\textwidth}
    \includegraphics[width=\textwidth]{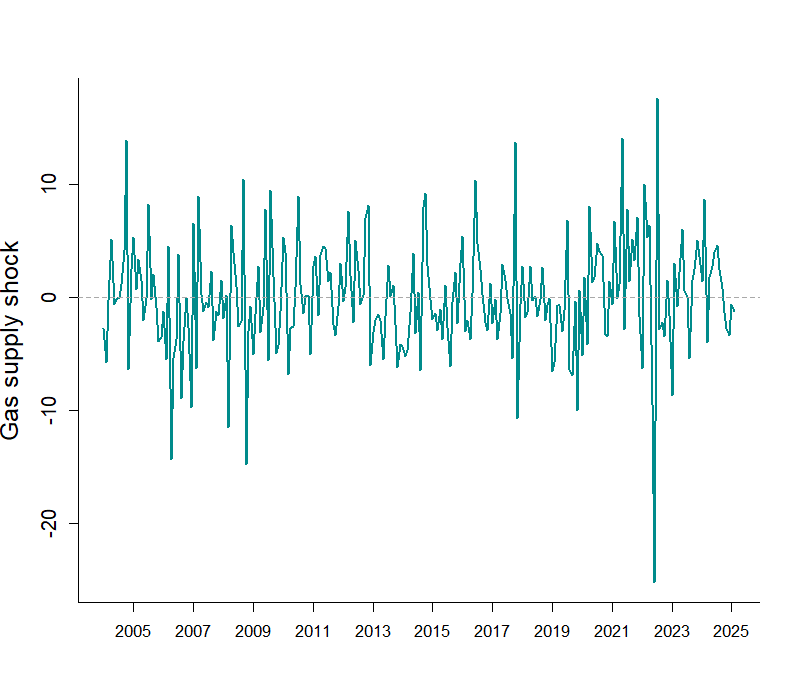}
    \end{subfigure}
    \begin{subfigure}[h]{0.4\textwidth}
    \includegraphics[width=\textwidth]{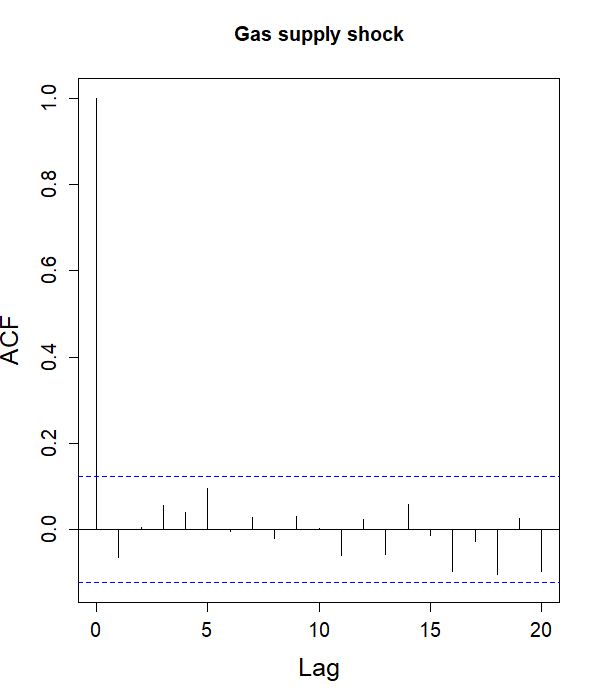}
    \end{subfigure}
    \end{minipage}
    \caption{\scriptsize \justifying \textbf{Gas supply shock.} The figure shows the natural gas supply shock obtained from the IV-VAR (left) and the corresponding correlogram (right). Sample: 2005:M4–2025:M6.}
    \label{figura 10}
\end{figure}

\noindent We aggregate the daily structural shock at monthly and quarterly frequencies by sum.\footnote{Our results are robust to take instead the monthly average.} The aggregated shock is reported in \autoref{figura 10} together with its correlogram.\footnote{In \Cref{app. hf responses}, we report the impulse response functions from the daily VAR-X. The identified shock generates a persistent increase in spot natural gas prices on impact, while responses of financial market variables and electricity demand are small and statistically insignificant, providing robustness against financial- and demand-driven contamination.} 
The daily VAR must be specified in such a way that it contains enough information to identify $\epsilon_{1,t}$, distinguishing it with respect to any other shock. Since the model does not contain low frequency macroeconomic variables, this is a concern and we should test wether the model contains enough information to identify the structural shock of interest. Hence, we run the invertibility test suggested in \cite{FORNI2014124}.

We regress the aggregated structural shock on up to 10 lagged factors extracted from the high-dimensional dataset for Italy constructed by \cite{barigozzi2024large}, as well as on one lag of the aggregated shock itself.\footnote{The number of factors is selected using the criterion of \cite{bai2002determining}. When the selected number exceeds 10, we project the shock onto the first 10 factors, following the recommendation of \cite{FORNI2014124}.} The results of the test are reported in \autoref{tab:invertibility_1}. We find no strong evidence against invertibility, as the F-statistic is not significant in either the monthly or the quarterly specification. This suggests that the aggregated shock lies in the information set spanned by the low-frequency model, and can therefore be used as a valid external instrument in the subsequent stages of the analysis.

\begin{table}[H]
\centering

\begin{tabular}{lcl|lcl}
\toprule
                     & $\textbf{Estimate}$ &  $\textbf{p value}$    &            & $\textbf{Estimate}$ &  $\textbf{p value}$ \\
\midrule
Intercept              &  0.033 &  0.915       & Intercept          &  0.095     &  0.927  \\
$F^M_{1,t-1}$          & -0.391 &  0.522       & $F^Q_{1,t-1}$      & -0.047     &  0.976  \\
$F^M_{2,t-1}$          &  0.014 &  0.981       & $F^Q_{2,t-1}$      &  1.388     &  0.496  \\
$F^M_{3,t-1}$          & -0.468 &  0.624       & $F^Q_{3,t-1}$      &  0.678     &  0.854  \\
$F^M_{4,t-1}$          &  1.324 &  0.175       & $F^Q_{4,t-1}$      & -0.907     &  0.783  \\
$F^M_{5,t-1}$          &  0.233 &  0.791       & $F^Q_{5,t-1}$      & -3.824     &  0.308  \\
$F^M_{6,t-1}$          & -1.722 &  0.194       & $F^Q_{6,t-1}$      &  4.75      &  0.263  \\
$F^M_{7,t-1}$          &  2.084 &  0.127       & $\epsilon_{1,t-1}$ & 0.062      &  0.636  \\
$F^M_{8,t-1}$          &  4.943 &  $0.023^{**}$&                    &            &         \\
$F^M_{9,t-1}$          &  0.806 &  0.765       &                    &            &         \\
$F^M_{10,t-1}$         &  2.330 &  0.122       &                    &            &         \\
$\epsilon_{1,t-1}$     & -0.076 &  0.329       &                    &            &         \\
\midrule
\textbf{F statistic}          &         & 1.528         & \textbf{F statistic}        &            & 0.371   \\
\textbf{p value}              &         & 0.13          & \textbf{p value}            &            & 0.895   \\

\bottomrule
\end{tabular}
\caption{\scriptsize \justifying \textbf{Invertibility test results.} The table reports the results of the \cite{FORNI2014124} invertibility test conducted at the monthly (left) and quarterly (right) frequency. The dependent variable is the lower frequency shock, obtained by aggregating the daily structural shock at the lower frequency. Regressors are one-period lags of the factors $F_{i,t}^f$ at the desider frequency $f$, and the one-period lag of the shock. Standard errors are Newey-West with 12 lags. Significance: $^{***}p<0.01$, $^{**}p<0.05$, $^{*}p<0.10$.}
\label{tab:invertibility_1}
\end{table}

\paragraph{Shock exogeneity} We assess whether the aggregated gas supply shock contains information already captured by other energy-related structural shocks identified in the literature. To this end, we perform Wald tests of joint significance by regressing the monthly gas supply shock on the contemporaneous and 12 lagged values of each proxy shock, as well as on 12 lags of the gas shock itself. We consider the oil price news shock of \cite{Kanzig2020}, the carbon price shock of \cite{kanzig2023unequal}, several oil demand and supply shocks from \cite{caldara2019oil}, \cite{BH2019}, \cite{kilian2008exogenous}, \cite{kilian2009not}, and the oil price expectation shock of \cite{Baumeister2023}. Results are reported in \autoref{tab:exogeneity_tests}. In all cases, the null hypothesis that the gas supply shock is not spanned by other energy-related innovations cannot be rejected, suggesting that it captures distinct information.

\begin{table}[H]

\centering
\scriptsize
\begin{tabular}{llcc}
  \toprule
   Innovation & Sample & Wald statistic & P-value \\ 
   \midrule
   \cite{Kanzig2020} Oil price news                            & 2005M4 - 2024M6  & 12.4996  & 0.4064  \\
   \cite{kanzig2023unequal} Carbon price                       & 2005M4 - 2019M12 & 17.3019  & 0.1386  \\ 
   \cite{caldara2019oil} Oil demand                            & 2005M4 - 2015M11 & 8.387    & 0.7542  \\
   \cite{caldara2019oil} Oil supply                            & 2005M4 - 2015M11 & 13.7724  & 0.3155  \\ 
   \cite{BH2019} Oil supply                                    & 2005M4 - 2025M5  &  8.3373  & 0.7582  \\
   \cite{BH2019} Oil consumption demand                        & 2005M4 - 2025M5  & 11.6629  & 0.4731  \\
   \cite{BH2019} Oil inventories                               & 2005M4 - 2025M5  & 13.4428  & 0.3377  \\
   \cite{Baumeister2023} Oil price expectation                 & 2005M4 - 2023M3  &  7.8767  &  0.7946  \\
   \cite{kilian2009not} Oil supply                             & 2005M4 - 2009M12 &  5.7017  & 0.9304  \\
   \cite{kilian2008exogenous} Oil Supply                       & 2005M4 - 2016M12 & 16.1255  & 0.1855  \\
   
  \bottomrule
\end{tabular}
\caption{\scriptsize \justifying \textbf{Wald test results.} The table reports Wald test statistics for the joint significance of coefficients. The null hypothesis is that all coefficients are zero, indicating no contamination.}
\label{tab:exogeneity_tests}
\end{table}

\begin{figure}[H]
  \centering

  \begin{minipage}{\linewidth}

    \centering
    \begin{subfigure}[t]{0.32\textwidth}
      \centering
      \includegraphics[width=\linewidth]{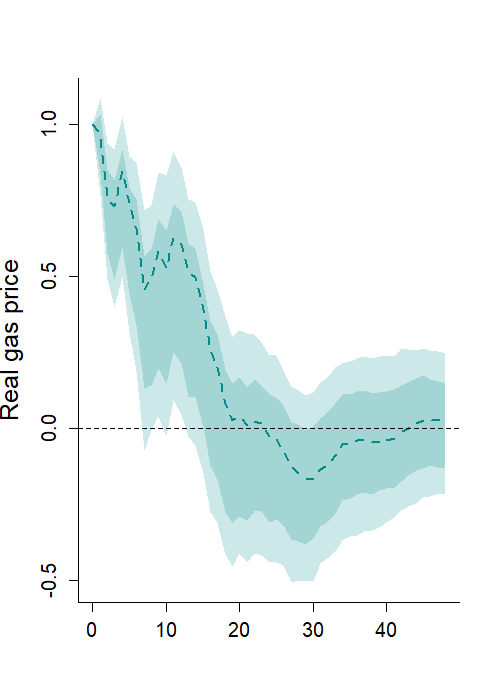}

    \end{subfigure}\hfill
    \begin{subfigure}[t]{0.32\textwidth}
      \centering
      \includegraphics[width=\linewidth]{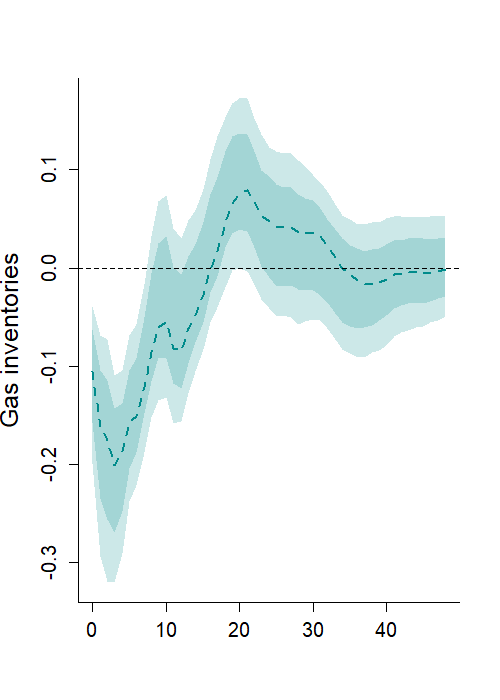}

    \end{subfigure}\hfill
    \begin{subfigure}[t]{0.32\textwidth}
      \centering
      \includegraphics[width=\linewidth]{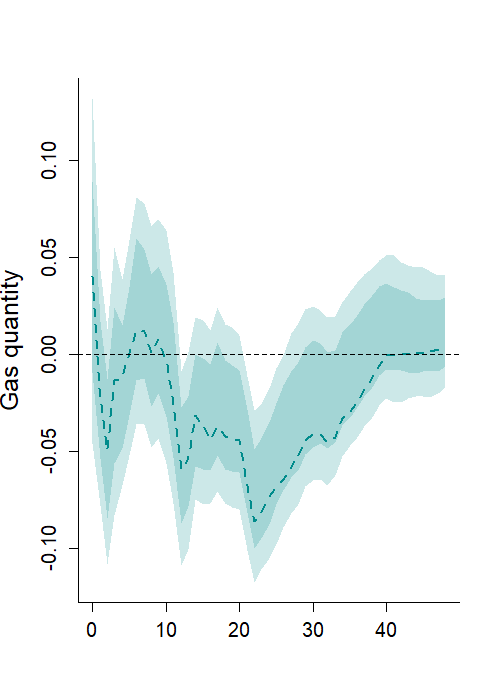}

    \end{subfigure}

    \vspace{0.2em}

    \begin{subfigure}[t]{0.32\textwidth}
      \centering
      \includegraphics[width=\linewidth]{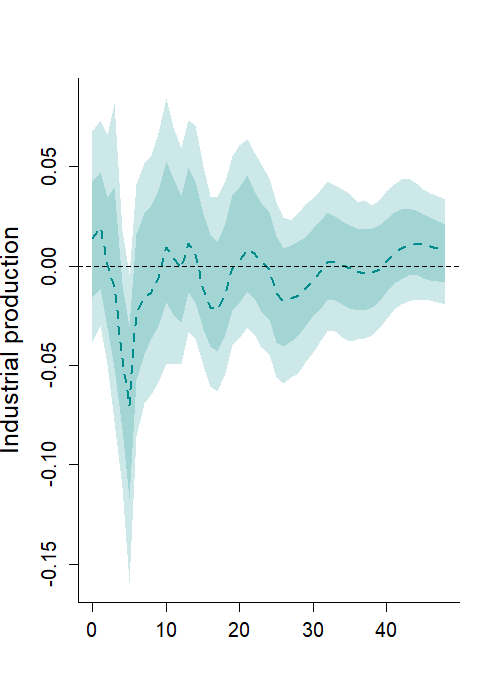}

    \end{subfigure}\hfill
    \begin{subfigure}[t]{0.32\textwidth}
      \centering
      \includegraphics[width=\linewidth]{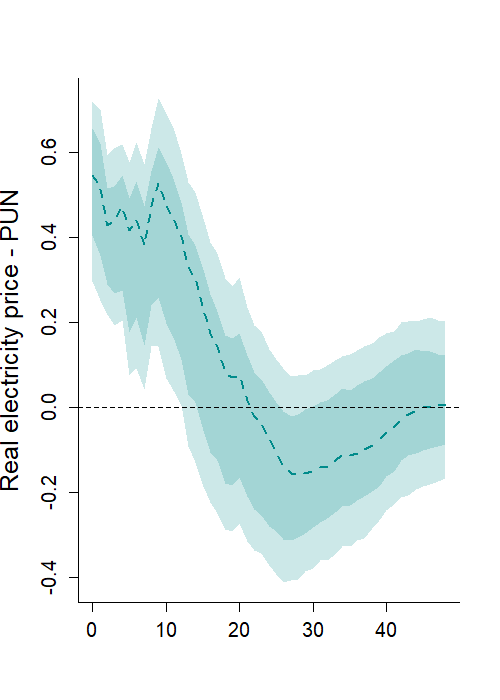}

    \end{subfigure}\hfill
    \begin{subfigure}[t]{0.32\textwidth}
      \centering
      \includegraphics[width=\linewidth]{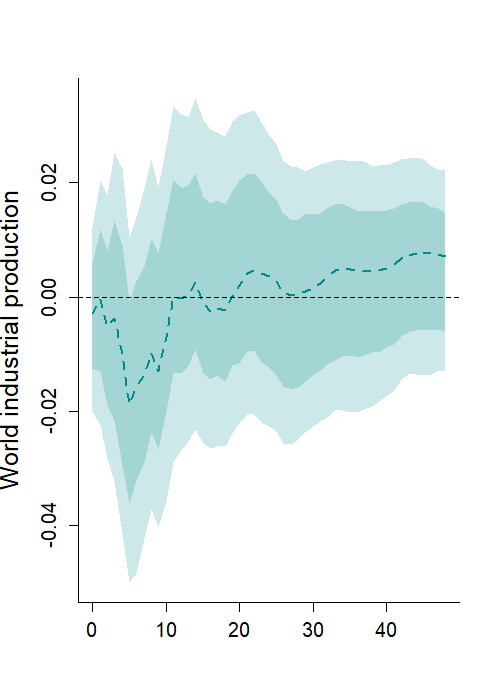}

    \end{subfigure}

    \caption{\scriptsize\justifying \textbf{Impulse response functions of gas market variables, electricity prices, and industrial production to a natural gas supply shock.}
    The figure reports impulse response functions of the spot real natural gas price, gas inventories, gas quantity, domestic industrial production, the real electricity price (PUN), and world industrial production to a natural gas supply shock. Shaded areas represent 68\% and 90\% confidence intervals computed using a moving block bootstrap. Source: Ministero dell’Ambiente e della Sicurezza Energetica; Gestore dei Mercati Energetici; ISTAT; \cite{BH2019}. Sample: 2004:M1--2025:M6.}
    \label{figura 11}
  \end{minipage}
\end{figure}

\paragraph{Monthly VAR} We specify a monthly VAR including six variables. We take the day ahead Dutch TTF as a measure of the spot price of natural gas. We make this price real by deflating it by a measure of the Italian HICP we built by excluding the price of natural gas. Then, in the spirit of \cite{adolfsen2024gas}, we include the available gas quantities, gas inventories and industrial production for Italy. We also control for the world industrial production and for the real "Prezzo Unico Nazionale" (PUN) as a measure of the Italian spot electricity price.\footnote{Data about gas quantities and gas inventories are provided by the Italian \href{https://sisen.mase.gov.it/dgsaie/bilancio-gas-naturale}{
Ministry of Environment and Energy Security}. Gas quantities are defined as national production, plus imports minus exports. Gas inventories are given by the cumulative sum of the variation in inventories, given a base value of 7000 millions of standard cubic meters. Data about the PUN are provided by \href{https://www.mercatoelettrico.org/en/Default.aspx}{Gestore Mercati Energetici} (GME). World industrial production is provided by \cite{BH2019}.} Both gas quantities gas inventories and the PUN are seasonally adjusted by using the X-13 ARIMA seasonal adjustment method. Monitoring the response of the PUN to a gas supply shock is informative because, in a uniform-price electricity market with gas-fired units often setting the marginal cost, like the Italian one, changes in the TTF day-ahead price feed into generators’ bids and hence the Italian spot price with short lags; the PUN therefore reflects downstream pass-through from gas to electricity price. 
The VAR is estimated in log levels with a constant from 2004M1 until 2025M6. The lag order is set at 12. 
We use the daily structural shock, aggregated at the monthly frequency, as an external instrument for the monthly VAR. The instrument satisfies the relevance condition, with an F-statistics of 74.38 and an heteroskedasticity robust F statistics of 55.71. Impulse response functions are reported in \autoref{figura 11}. As in \cite{kanzig2021macroeconomic}, we can interpret the impulse response functions as elasticities, since the model is estimated in log levels. We normalize the shock to increase the spot natural gas price  by 1\%.

The TTF responds positively, and the response reverts to 0 in about one year. The increase in the price of natural gas is then transmitted to the national electricity price, measured by the PUN. Gas-fired plants set the marginal price, so the electricity market mirrors gas dynamics with a short lag. About half of the increase in the gas price is transmitted to the electricity price. Gas inventories decrease by 0.1\%. Higher gas prices coincide with withdrawals from storage, before restocking once prices ease. 
Gas quantity does not respond to the shock on impact. After 10 months, we observe a small but persistent drop followed by a gradual recovery. The effect on the domestic industrial production are lagged and short lived. Few months after the shock we observe a quick drop of 0.05\% in industrial production, which quickly reverts to 0. Hence, consistently with the transitory nature of the shock and with the findings of \cite{guntner2024sudden}, a gas supply has limited effects on industrial production.\footnote{To verify the robustness of our results, we estimate the model including the measure of geopolitical risk by \cite{caldara2022measuring}. Adding this additional variable to the model does not affect the strength of our instrument. The F statistic is 68.49 and the heteroskedasticity robust F statistic is instead 60.539. Impulse response functions for the model that includes the measure of geopolitical risk are reported in \cref{App. IVrobust} of the Appendix. Our baseline results are not affected by the inclusion in the model of this additional variable. This evidence suggests that our instrument is not merely capturing a generic geopolitical risk shock, but predominantly identifies shocks to the physical supply of natural gas relevant for Italy.}

\paragraph{Quarterly VAR: Natural gas supply shocks and firms' behavior} We take the daily shock, aggregated at the quarterly frequency, and we use it as an external instrument in a VAR estimated from 2005Q1 until 2025Q2. The model contains the real spot natural gas price, unemployment rate, the nominal interest rate and the log HICP and industrial production index. Following \cite{kanzig2021macroeconomic}, we estimate it several times, each time adding as an additional last variable one of the aggregate measures for firms we extract from the SIGE survey. 

\begin{figure}[H]
    \centering
    \begin{minipage}{1\linewidth}
    \centering
    
    \begin{subfigure}[t]{0.32\textwidth}
      \centering
      \includegraphics[width=\linewidth]{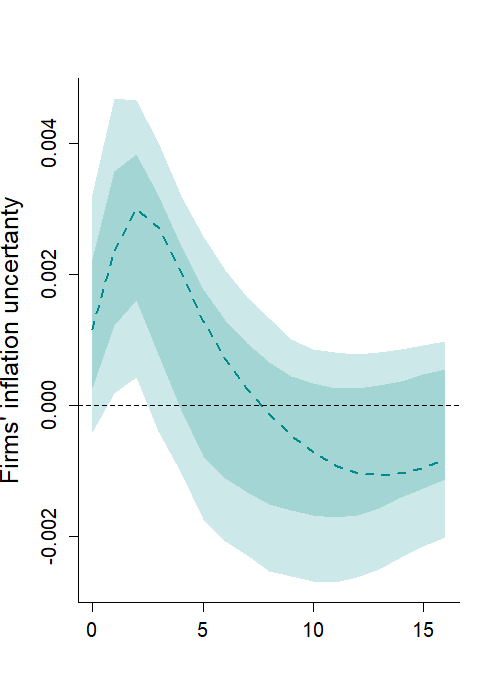}
    \end{subfigure}\hfill
    \begin{subfigure}[t]{0.32\textwidth}
      \centering
      \includegraphics[width=\linewidth]{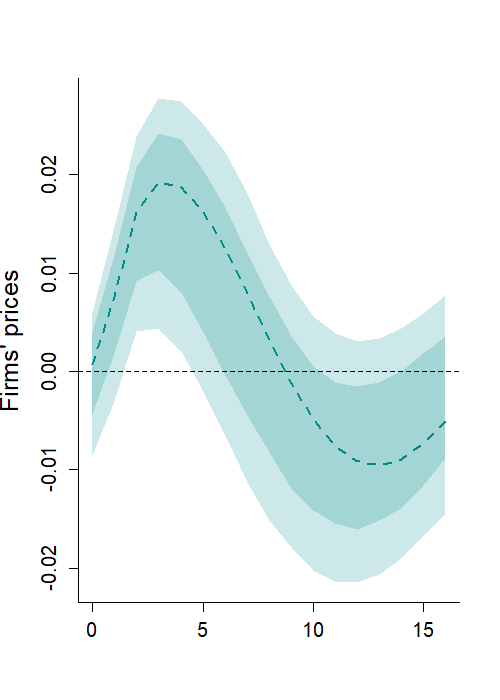}
    \end{subfigure}\hfill
    \begin{subfigure}[t]{0.32\textwidth}
      \centering
      \includegraphics[width=\linewidth]{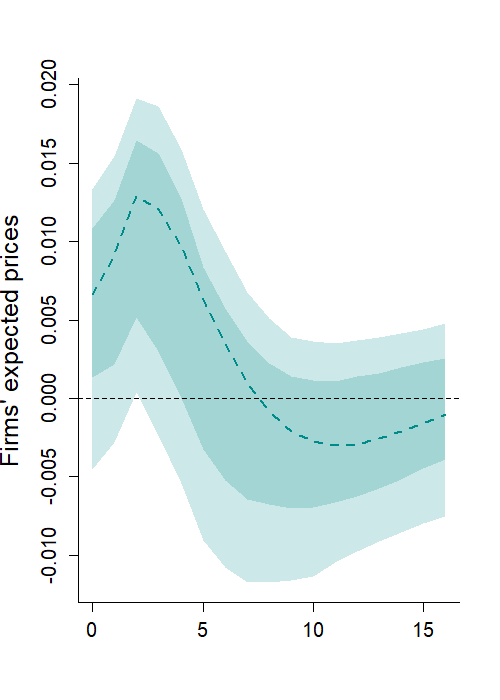}
    \end{subfigure}

    \vspace{0.2em}

    \begin{subfigure}[t]{0.32\textwidth}
      \centering
      \includegraphics[width=\linewidth]{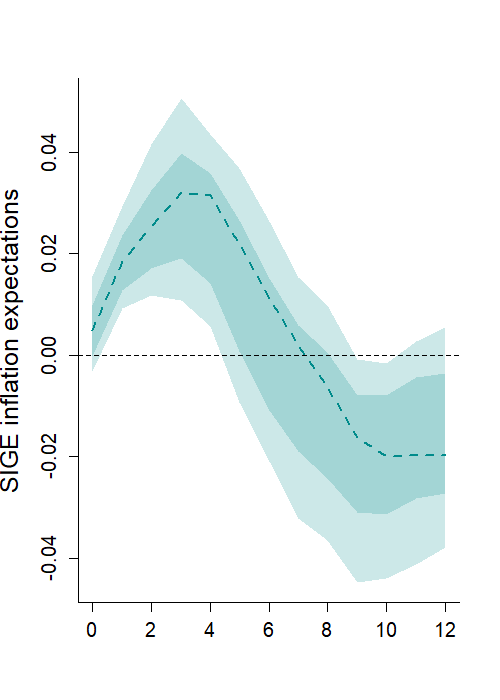}
    \end{subfigure}\hfill
    \begin{subfigure}[t]{0.32\textwidth}
      \centering
      \includegraphics[width=\linewidth]{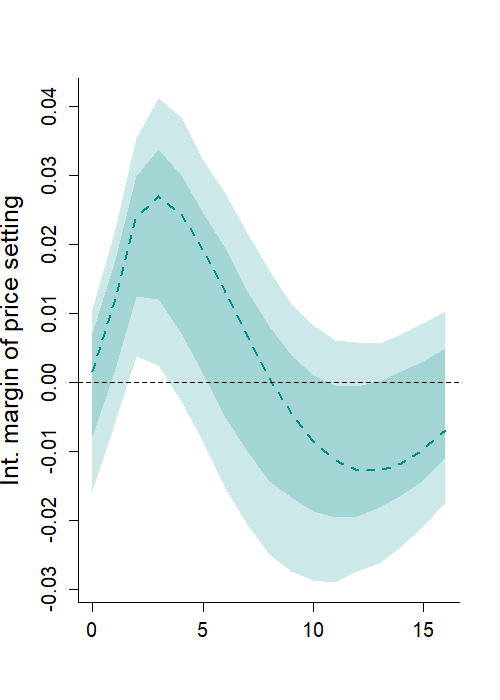}
    \end{subfigure}\hfill
    \begin{subfigure}[t]{0.32\textwidth}
      \centering
      \includegraphics[width=\linewidth]{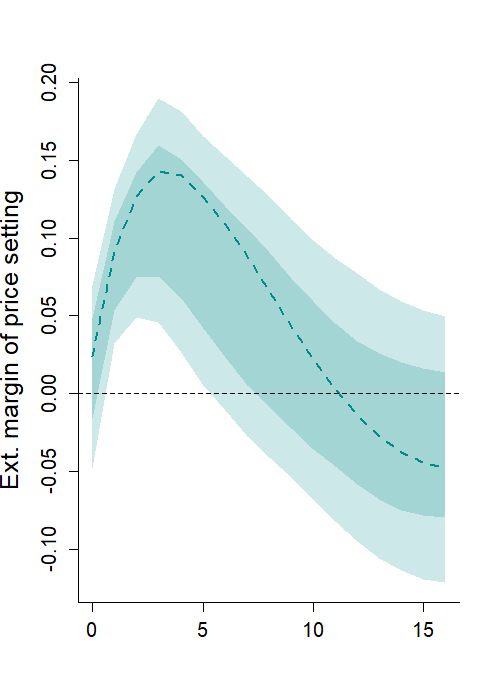}
    \end{subfigure}
    
    \caption{\scriptsize \justifying \textbf{Impulse response functions of firms' related variables to a natural gas supply shock.} The figure reports the impulse response functions of the inflation uncertainty index, firms' realized and expected annual price change, inflation expectations, intensive and extensive margins to a natural gas supply shock. Shaded areas represent 68\% and 90\% confidence intervals computed through a moving block bootstrap procedure. Source: Bank of Italy, Survey on inflation and growth expectations. Sample: 2005:Q1–2025:Q2.}
    \label{figura 5A}
    \end{minipage}
\end{figure}

Specifically, we consider firms’ inflation uncertainty, computed by using the \cite{Jurado2015} approach and by exploiting the large dataset for the Italian Economy provided by \cite{barigozzi2024large}, as well as expected and realized prices, inflation expectations, and the intensive and extensive margins of price setting.
We set the number of lags to 2 given the short sample length, which requires a parsimonious specification to preserve degrees of freedom while still capturing short-run dynamics.
After checking that the condition for instrument relevance is satisfied for each model specification, we simulate a shock that increases the real price of natural gas by 1\% and we compute the impulse response functions, reported in \autoref{figura 5A}. The results are consistent with a cost-push disturbance. Inflation uncertainty increases, suggesting that the shock not only pushes up the level of prices but also complicates firms’ forecasting.\footnote{As we show in \Cref{App. IVrobust} of the Appendix, this result is robust to using the \cite{binder2017measuring} measure of firms' inflation uncertainty we estimate from the round responses in the SIGE.}

Firms’ prices exhibit a hump-shaped response, increasing persistently. Their response reverts to 0 in 5 quarters. 
The pass-through of input cost shocks to firms’ selling prices is however quantitatively limited. A gas supply shock that raises the real natural gas price by 1\% translates into a peak increase of about 0.02 percentage points in firms’ annual price inflation. 

One-year-ahead expectations about future prices display a shorter-lived increase: on impact, they increase by about 0.005 percentage points and revert to 0 within half a year. However, their response is significant just at the 68\% confidence level.  Firms’ one-year-ahead inflation expectations increase strongly and persistently, showing a hump-shaped response. The response of the price margins indicates that the shock not only increases the proportion of firms that update their price by 0.05\%; it also induces the price-adjusting firms to adjust more aggressively when they do update, as the intensive and extensive margins rise gradually. 

\paragraph{Sectoral analysis} \cite{riggi2022price} document that firm size and geographical location do not significantly affect the magnitude of pass-through following an input price shock. In contrast, they find that sectoral affiliation plays a relevant role. To quantify differences in the pass-through of a gas supply shock across sectors, we re-estimate the models described in the previous paragraph separately for the manufacturing and services sectors, considering expected prices and inflation, price changes over the past year, and both the intensive and extensive margins of price adjustment.\footnote{As no significant differences emerge in terms of uncertainty among these two sector, we use the aggregate index of uncertainty for the total of the firms.}

\begin{figure}[H]
  \centering

  \begin{minipage}{\linewidth}
    \centering

    \begin{subfigure}[t]{0.32\textwidth}
      \centering
      \includegraphics[width=\linewidth]{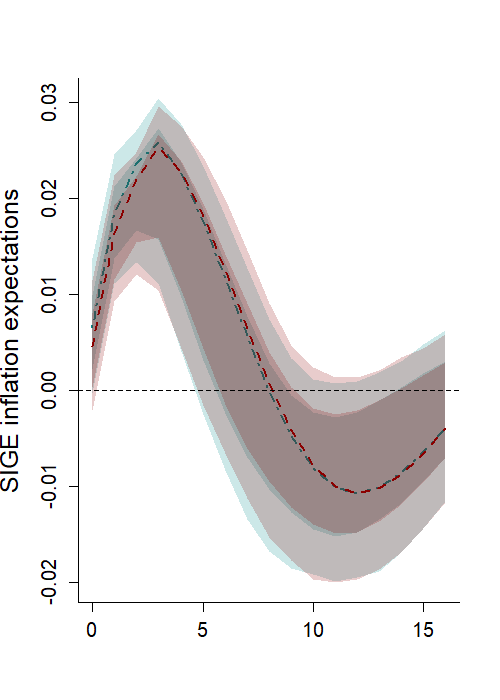}
    \end{subfigure}\hfill
    \begin{subfigure}[t]{0.32\textwidth}
      \centering
      \includegraphics[width=\linewidth]{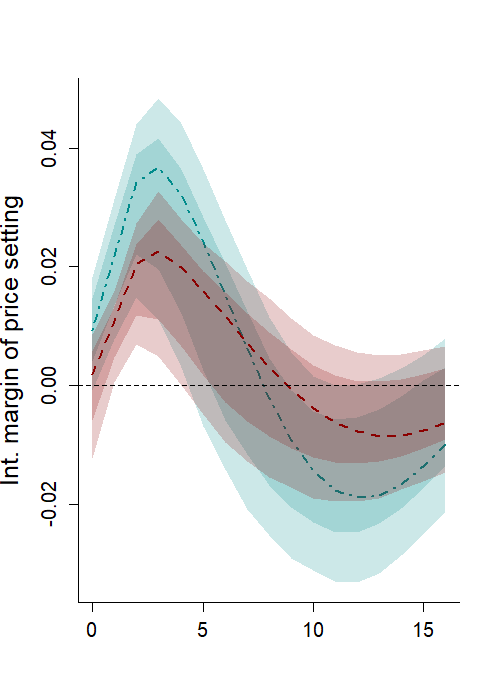}
    \end{subfigure}\hfill
    \begin{subfigure}[t]{0.32\textwidth}
      \centering
      \includegraphics[width=\linewidth]{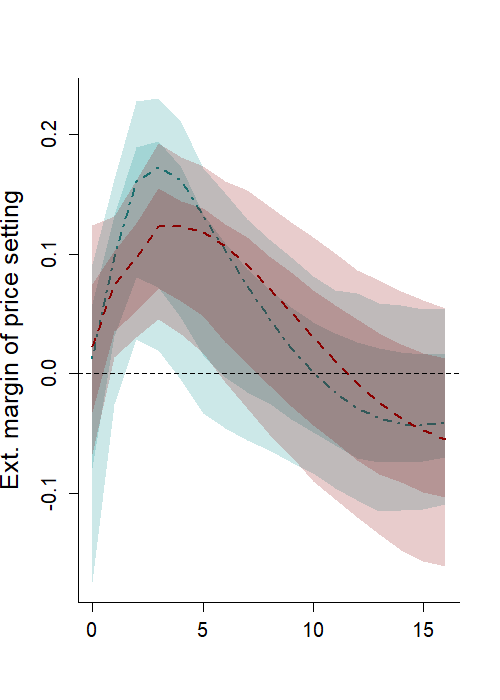}
    \end{subfigure}

    \vspace{0.6em}

    \begin{subfigure}[t]{0.32\textwidth}
      \centering
      \includegraphics[width=\linewidth]{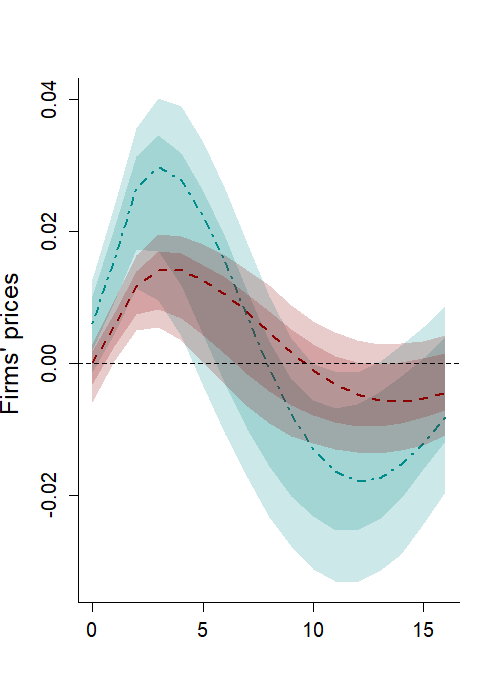}
    \end{subfigure}\hfill
    \begin{subfigure}[t]{0.32\textwidth}
      \centering
      \includegraphics[width=\linewidth]{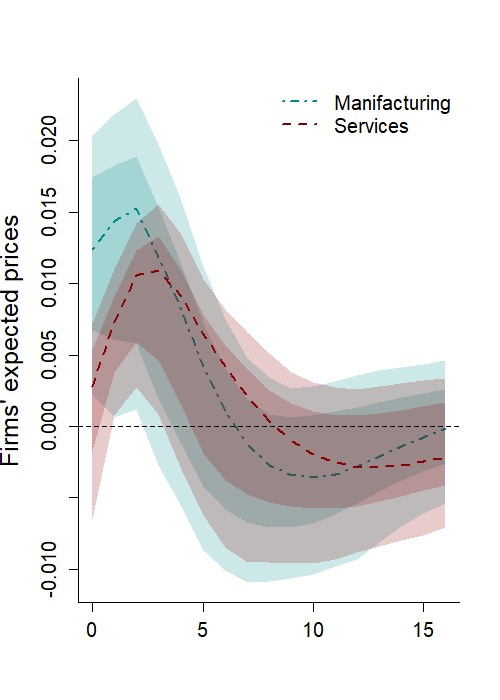}
    \end{subfigure}\hfill
    \begin{subfigure}[t]{0.32\textwidth}
      \centering
      \rule{\linewidth}{0pt}
    \end{subfigure}

    \caption{\scriptsize\justifying \textbf{Differences in the IRFs across sectors.}
    The figure reports the impulse response functions of firms’ inflation expectations, intensive and extensive margins, and expected and realized annual price changes in response to a natural gas supply shock, separately for manufacturing firms (light blue) and service-sector firms (red). Shaded areas represent 68\% and 90\% confidence intervals computed using a moving block bootstrap procedure. Source: Bank of Italy, Survey on Inflation and Growth Expectations. Sample period: 2005:Q1--2025:Q2.
    }
    \label{figura 5E}
  \end{minipage}
\end{figure}

This allows us to analyze differences in firms’ price-setting behavior across the two sectors conditional on a natural gas supply shock. \autoref{figura 5E} reports the estimated impulse response functions of the price-setting variables for manufacturing and services firms.
For manufacturing firms, the shock induces a markedly stronger response across all price-setting dimensions. Both realized and expected prices increase more sharply and reach their peak earlier than in the services sector, indicating a more aggressive adjustment to higher input costs. This is consistent with manufacturing firms being more energy-intensive than service-sector firms. 
The intensive margin exhibits a similarly pronounced reaction, with manufacturing firms increasing the size of their price changes more strongly in the short run. The extensive margin also shows a significant expansion, suggesting that a larger share of manufacturing firms revises prices following the shock. 
In contrast, while the response of service-sector firms along the extensive margin is comparable in magnitude, it appears to be more persistent over time.

Overall, the results indicate that manufacturing firms react more promptly and forcefully to natural gas supply shocks, particularly in terms of the size of price adjustments, while the likelihood of adjusting prices responds similarly across the two sectors. These differences arise despite the fact that the impulse response functions from the two sector-specific models do not display meaningful differences in the response of inflation expectations to the shock. This finding is consistent with \cite{riggi2022price}, who show that inflation expectations are not a key determinant of firms’ price-setting behavior, which is instead more strongly influenced by firms’ expectations about the future path of their own prices.

\subsection{State dependent local projections}

In this section, we assess the presence of nonlinearities in the transmission of a natural gas supply shock on firms' price setting behavior. In their analysis for the Euro Area, \cite{adolfsen2024gas} find that natural gas supply shocks are characterized by a non linear transmission with  respect to labor market tightness, with effects that are considerably magnified when the economy operates near full capacity. Starting from this evidence, we show that the pass-through of natural gas supply shocks is also characterized by substantial nonlinearities related to the level of uncertainty surrounding future inflation by recurring to state dependent local projections \citep{ramey2018government, F2021}.
To determine regimes of high and low inflation uncertainty, we take as a baseline the measure of inflation uncertainty derived by adopting the methodology of \cite{Jurado2015}.\footnote{As we show in \Cref{App. sdlp} of the Appendix, results are robust to using as an alternative the measure of \cite{binder2017measuring}, and the impulse response functions we obtain are essentially identical.} 

Following \cite{F2021} we scale the uncertainty index for the average inflation expectations of firms, in order to account for high inflationary periods, and we smooth it by applying a 8 period backward looking weighted moving average filter. 
We assume that the transition between states of high and low probabilities of inflation uncertainty is governed a logistic function $Z(\hat{\Delta}_{t-1})\in[0,1]$ of the lagged index, defined as:

\begin{equation}
    Z(\hat{\Delta}_{t-1}) = \frac{exp\left(\eta\frac{\hat{\Delta}_{t-1} - \mu}{\sigma_{\hat{\Delta}}}\right)}{1 + exp\left(\eta\frac{\hat{\Delta}_{t-1} - \mu}{\sigma_{\hat{\Delta}}}\right)}
    \label{equazione 2}
\end{equation}

\noindent where $\hat{\Delta}_t$ is the state variable, $\eta$ is a parameter that determines the steepness of the transition, and $\mu$ and $\sigma_{\hat{\Delta}}$ represent the median and standard deviation of the state variable, respectively. The estimation of the parameter $\eta$ is challenging due to the nonlinearity of the logistic function and the strong identifying assumptions it entails. Following the existing literature, we therefore fix $\eta$ at 5. Our results are robust to a wide range of alternative values, with $\eta$ varying between 1 and 100.

\begin{figure}[H]
    \centering
    \begin{minipage}{1\linewidth}
    \begin{subfigure}[h]{1\textwidth}
    \includegraphics[width=\textwidth]{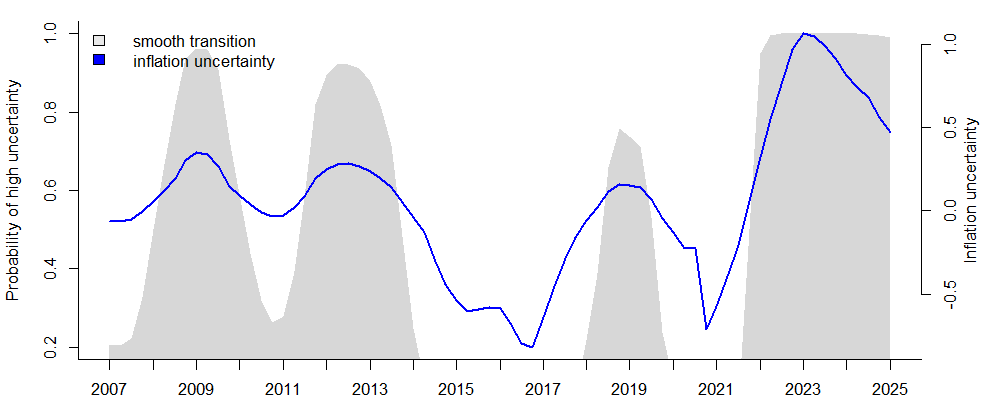}
    \end{subfigure}

    \caption{\scriptsize \justifying \textbf{Inflation uncertainty.} 
    Estimated probability of high inflation uncertainty following \cite{Jurado2015}, computed via smooth transition logistic function (shaded areas). Blue line represent the underlying state variable.
    Source: Bank of Italy, Survey on inflation and growth expectations. Sample: 2007:Q1–2025:Q2.}
    \label{figura 6}
    \end{minipage}
\end{figure}

The transition probabilities are reported in \autoref{figura 6}. The probability of being in the high uncertainty state increases during the onset of the GFC and remains high during the sovereign debt crisis. It increases during 2019, before falling again during the pandemic. Then it surges again after 2021 and remains high during the European energy crisis, until the end of our sample. Having determined the smooth transition probability across states, we estimate the following set of state dependent local projections:\footnote{ \cite{gonccalves2024state} raise concerns about potential bias in impulse response functions estimated using this methodology when the state variable is endogenous to the innovation. The bias increases with the ratio between the shock magnitude and the standard deviation of the state variable. In our application, we consider a one–standard-deviation shock, which raises the real natural gas price by approximately 10\%, a magnitude that is both realistic and economically relevant. At this scale, the marginal conditional response is an appropriate and sufficient object of analysis.} 

\begin{equation}
\footnotesize
\begin{aligned}
x_{t+h} = \; & \bigg[ \tilde{\beta}^Z(h)w^{gas}_t + \Gamma^Z_I I + \Gamma^Z_X X  \bigg] Z(\hat{\Delta}_{t-1}) + \bigg[ \tilde{\beta}^{\bar{Z}}_{x}(h)w^{gas}_t + \Gamma^{\bar{Z}}_I I + \Gamma^{\bar{Z}}_X X \bigg] \bar{Z}(\hat{\Delta}_{t-1}) 
+ \gamma_D D_t + \epsilon_{t+h}
\label{eq4bis}
\end{aligned}
\end{equation}

\noindent Where $Z(\hat{\Delta}_{t-1})$ denotes the probability of being in the high inflation uncertainty state, $\bar{Z} \equiv 1 - Z(\hat{\Delta}_{t-1})$, and $w_t^{\text{gas}}$ is the structural gas supply shock extracted from the quarterly VAR identified in \Cref{sec:BPSVAR}. $X_t$ includes control variables, namely lagged values of the shock and of the dependent variable. The matrix $I_t$ collects interaction terms between the shock and (i) labor market tightness, (ii) a dummy equal to one when the annual HICP inflation rate exceeds 2\%, and (iii) a dummy equal to one when the economy is constrained at the zero lower bound. By controlling for these interactions in each state, we aim to isolate the nonlinear effects associated with inflation uncertainty from other well-documented sources of nonlinearity in the pass-through of supply shocks. In particular, controlling for the interaction between the ZLB dummy and the shock allows us to correctly separate the low uncertainty state from the monetary policy regime. 
Additionally, we control for the lags of these interactions. Finally, $D_t$ is a dummy variable that assumes value of 1 in correspondence of 2020Q2 and then decays exponentially over the subsequent quarters.\footnote{Removing this dummy from our specification does not affect our results.} 

\begin{figure}[H]
\centering

\parbox{\linewidth}{\centering\footnotesize\textit{Shock coefficient $\tilde{\beta}_h$ by inflation-uncertainty state }}\\
\begin{subfigure}{0.32\linewidth}\centering
    \includegraphics[width=\linewidth,keepaspectratio]{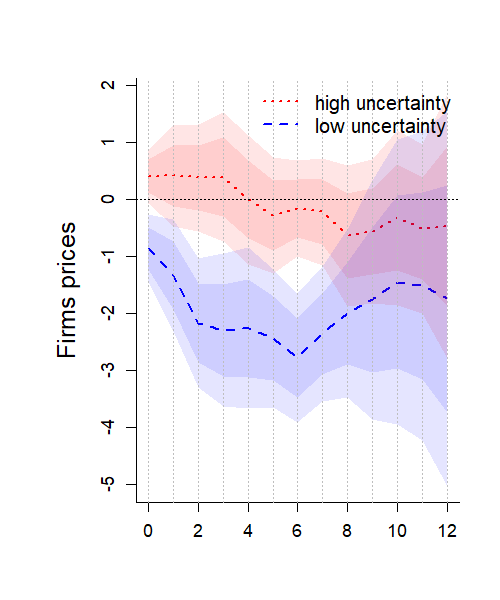}
\end{subfigure}
\begin{subfigure}{0.32\linewidth}\centering
    \includegraphics[width=\linewidth,keepaspectratio]{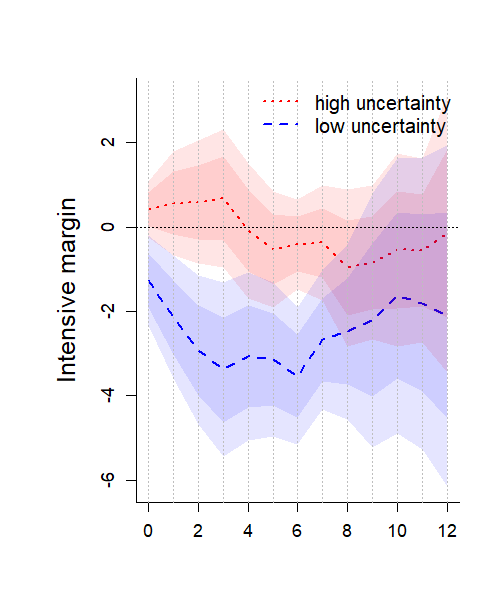}
\end{subfigure}
\begin{subfigure}{0.32\linewidth}\centering
    \includegraphics[width=\linewidth,keepaspectratio]{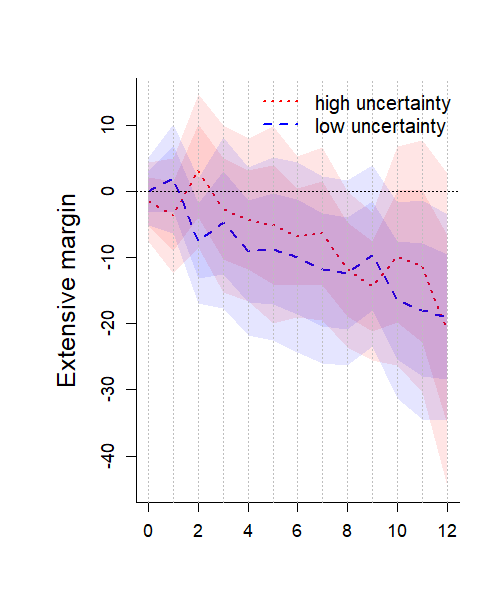}
\end{subfigure}\\[-0.25em]

\parbox{\linewidth}{\centering\footnotesize\textit{Interaction coefficient $\tilde{\gamma}_h$ (shock $\times$ labor market tightness) by inflation-uncertainty state}}\\
\begin{subfigure}{0.32\linewidth}\centering
    \includegraphics[width=\linewidth,keepaspectratio]{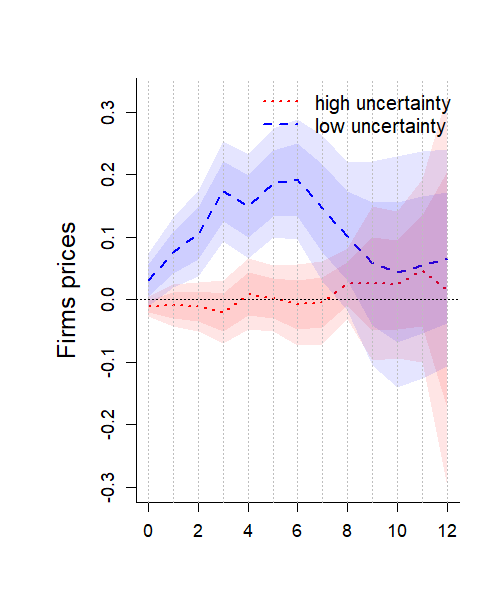}
\end{subfigure}
\begin{subfigure}{0.32\linewidth}\centering
    \includegraphics[width=\linewidth,keepaspectratio]{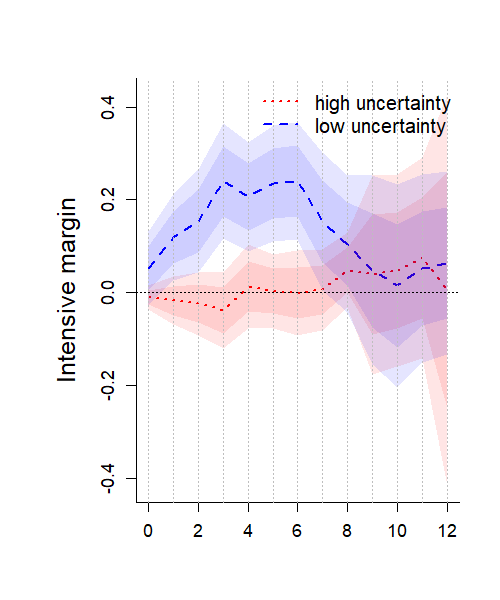}
\end{subfigure}
\begin{subfigure}{0.32\linewidth}\centering
    \includegraphics[width=\linewidth,keepaspectratio]{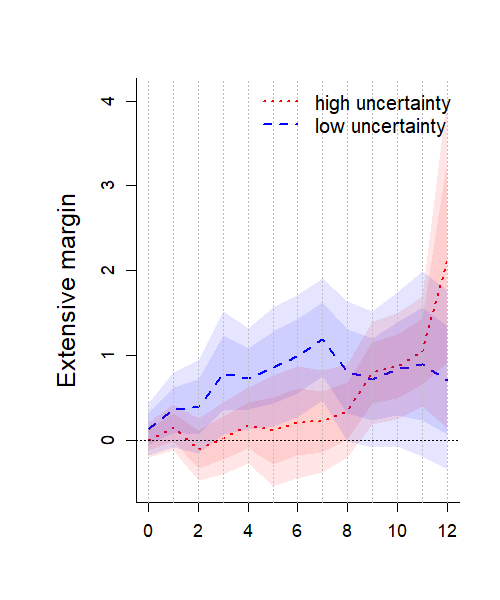}
\end{subfigure}\\[-0.25em]

\parbox{\linewidth}{\centering\footnotesize\textit{Interaction coefficient $\tilde{\delta}_h$ (shock $\times$ ZLB dummy) by inflation-uncertainty state}}\\
\begin{subfigure}{0.32\linewidth}\centering
    \includegraphics[width=\linewidth,keepaspectratio]{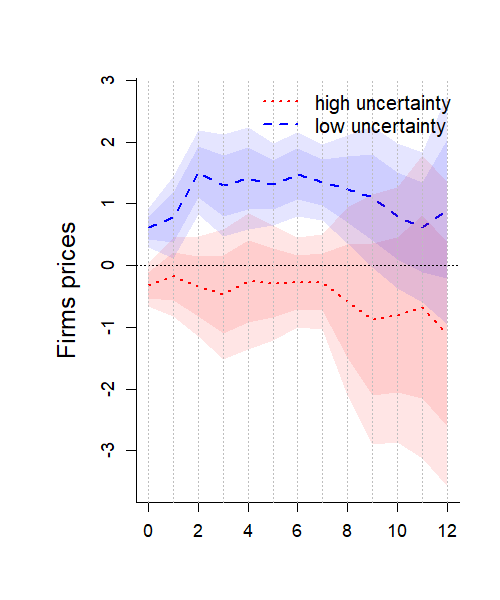}
\end{subfigure}
\begin{subfigure}{0.32\linewidth}\centering
    \includegraphics[width=\linewidth,keepaspectratio]{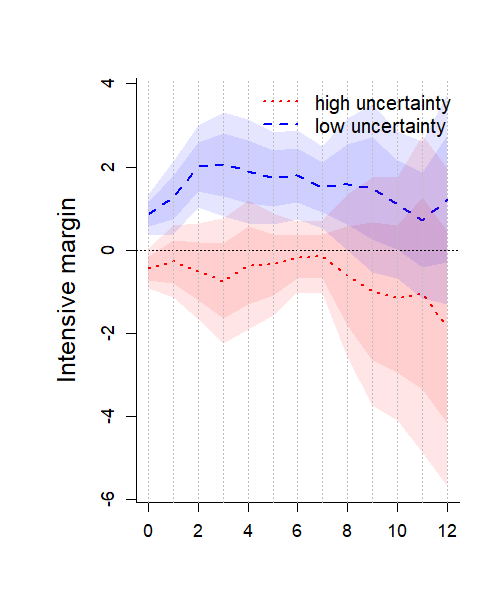}
\end{subfigure}
\begin{subfigure}{0.32\linewidth}\centering
    \includegraphics[width=\linewidth,keepaspectratio]{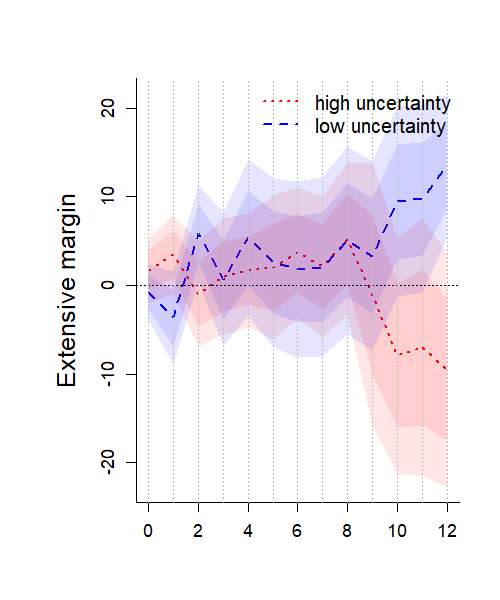}
\end{subfigure}

    \caption{\scriptsize \justifying \textbf{State dependent response of firms' prices and price setting margin.} 
    The figure shows the state-dependent responses of firms’ prices (left column) and the intensive (center column) and extensive (right column) price-setting margins to a natural gas supply shock. The first row reports the estimated coefficient on the shock, $\tilde{\beta}_h$, in high (red) and low (blue) inflation-uncertainty states at each horizon $h$. The second row reports the coefficient on the interaction between the shock and labor market tightness in each uncertainty state. The third row reports the coefficient on the interaction between the shock and the zero lower bound (ZLB) dummy in each uncertainty state. States of high and low inflation uncertainty are built starting from the measure of \cite{Jurado2015}. Bands denote 68\% and 90\% confidence intervals based on Newey--West standard errors. Source: Bank of Italy, Survey on Inflation and Growth Expectations; ISTAT, EUROSTAT.\emph{Sample: 2007Q1–2025Q2.}}

    \label{figura 7}
\end{figure}

\autoref{figura 7} reports the estimated impulse response functions. Under low uncertainty, firms’ prices decline in response to a natural gas supply shock. Firms anticipate the recessionary effects of the shock and, in order to preserve market shares, reduce their prices. This negative price response in the low-uncertainty state is primarily driven by the intensive margin of price setting, while we do not observe meaningful nonlinearities in the response of the extensive margin across the two states.

Consistent with \cite{adolfsen2024gas}, we find that a tight labor market is associated with a larger pass-through of the shock. However, this channel operates only in the low-uncertainty state. Similarly, we find qualitatively comparable but stronger effects when the economy is constrained at the zero lower bound. By contrast, we do not find evidence of nonlinearities associated with inflation being above 2\%.

\newpage
\section{Conclusions} \label{sec:4}

This paper provides new evidence on how natural gas supply shocks affect firms' pricing decisions and inflation expectations in Italy, a country heavily reliant on natural gas as its primary energy source. We identify a gas supply shock through a B-PSVAR framework and we analyze its implications for firms' pricing using quarterly survey data from the Bank of Italy's SIGE, spanning over two decades. Our analysis reveals three main findings.
First, natural gas supply shocks emerge as a significant driver of electricity prices in Italy. An 1\% increase in the spot natural gas price measured by the Dutch TTF future determines a 0.5\% increase in the Italian electricity price, measured by the PUN. 
Second, firms respond to natural gas supply innovations by adjusting both their current and expected own prices, with inflation uncertainty rising with a lag following the shock. These effects take about one year to dissipate fully, consistently with the transitory nature of energy price shocks documented in the literature. Hence, the increase of inflation uncertainty in response to the shock suggests that energy price disruptions affect not only the level of prices but also firms' ability in forecasting future inflation. Additionally, firms respond to the shock adjusting their prices more aggressively. Additionally, substantial differences emerge across sectors, with adjusting their prices more aggressively with respect to services sector firms. 
Third, we document substantial nonlinearities in firms' responses that depend on the pre-existing level of inflation uncertainty. When uncertainty is low, the recessionary effects of the shock dominate, causing firms to reduce prices below baseline as they anticipate that weakened demand conditions prevent cost pass-through.



\newpage
\begingroup
\setstretch{0.3}
\small
\bibliography{mybib}
\endgroup

\small

\pagebreak
\appendix
\appendixpage
\pagenumbering{Roman}
\renewcommand\thefigure{\thesection.\arabic{figure}}  
\renewcommand{\thetable}{\thesection\arabic{table}}
\renewcommand{\theequation}{\thesection.\arabic{equation}}   
\setcounter{figure}{0} 
\setcounter{table}{0}

\section{Factors driving firms' expectations about future prices}
\begin{figure}[H]
  
    \centering
    \begin{minipage}{0.9\linewidth}
    \begin{subfigure}[h]{1\textwidth}
    \includegraphics[width=\textwidth]{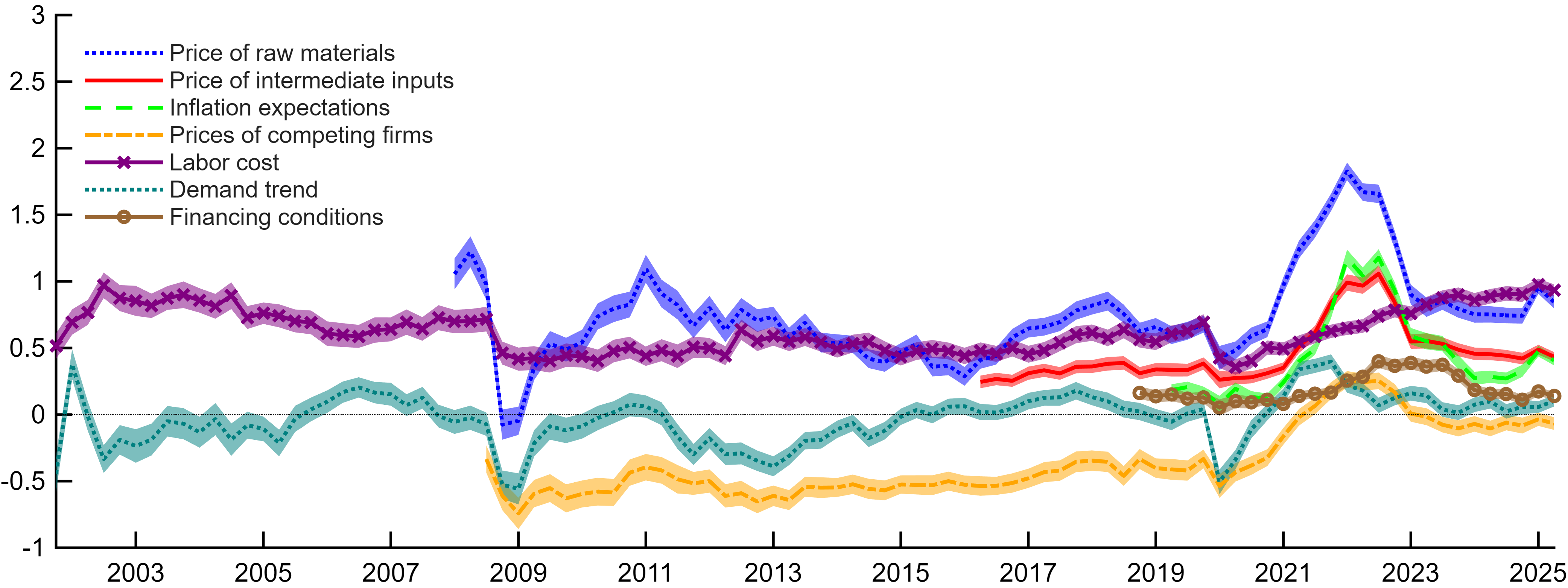}
    \end{subfigure}
       \caption{\scriptsize \justifying \textbf{Diffusion indexes on factors affecting firms' future prices.} The figure reports diffusion indexes computed on the categorical questions in the SIGE survey about the factors that can affect a firm's prices in the next year, together with 95\% confidence bands. Source: Survey of Inflation and Growth Expectation. Source: Bank of Italy, Survey on inflation and growth expectations. Sample: 1999:Q4–2025:Q2.}
    \label{figura 1}
    \end{minipage}
\end{figure}

To determine which factors are most likely to impact firms' price-setting behavior, we examine their responses to the categorical survey questions. Firms are asked about both the direction (positive, negative) and the intensity (strong, medium, modest) with which a specific factor is likely to affect their prices in the next year. The factors that are considered are the prices of raw materials and intermediate inputs, their inflation expectations for the next year, the prices of other competing firms, the trend in labor cost and aggregate demand and changes in the situation related to financing conditions. By using these responses, we build diffusion indexes using the approach of \cite{PintoSarteSharp2020}. Those indexes, shown in \autoref{figura 1} together with 95\% confidence bands, capture the average perceived intensity and direction of each factor across firms and are informative about movements in the distribution of firms' responses.\footnote{In order to build the indexes, each response is weighted by its intensity on a -3 to +3 scale, where higher absolute values indicate stronger perceived pressure.}

The evidence points to a clear cost-push narrative during the European energy crisis. 
The index for raw-material prices displays a consistently positive contribution over the entire sample, but it surges during the energy crisis, almost reaching the upper bound of the scale. This indicates an almost universal perception among firms that rising input costs were exerting strong upward pressure on their selling prices. Inflation expectations became the second most relevant upward pricing factor only around 2021, when the importance of intermediate-input costs also intensified. This pattern suggests that as firms observed widespread price increases, expectations began to feed directly into their own pricing decisions. 
Given the prominence of input costs—particularly energy-related raw materials—as key drivers of inflation during the post pandemic recovery, this motivates our focus on examining the effects of natural gas supply shocks on firm-level outcomes.

\section{Uncertainty index}\label{App. Uncertainty}

\subsection{Inflation uncertainty across updated and non-updated firms}

\begin{figure}[H]
  
    \centering
    \begin{minipage}{1\linewidth}
    \begin{subfigure}[h]{1\textwidth}
    \includegraphics[width=\textwidth]{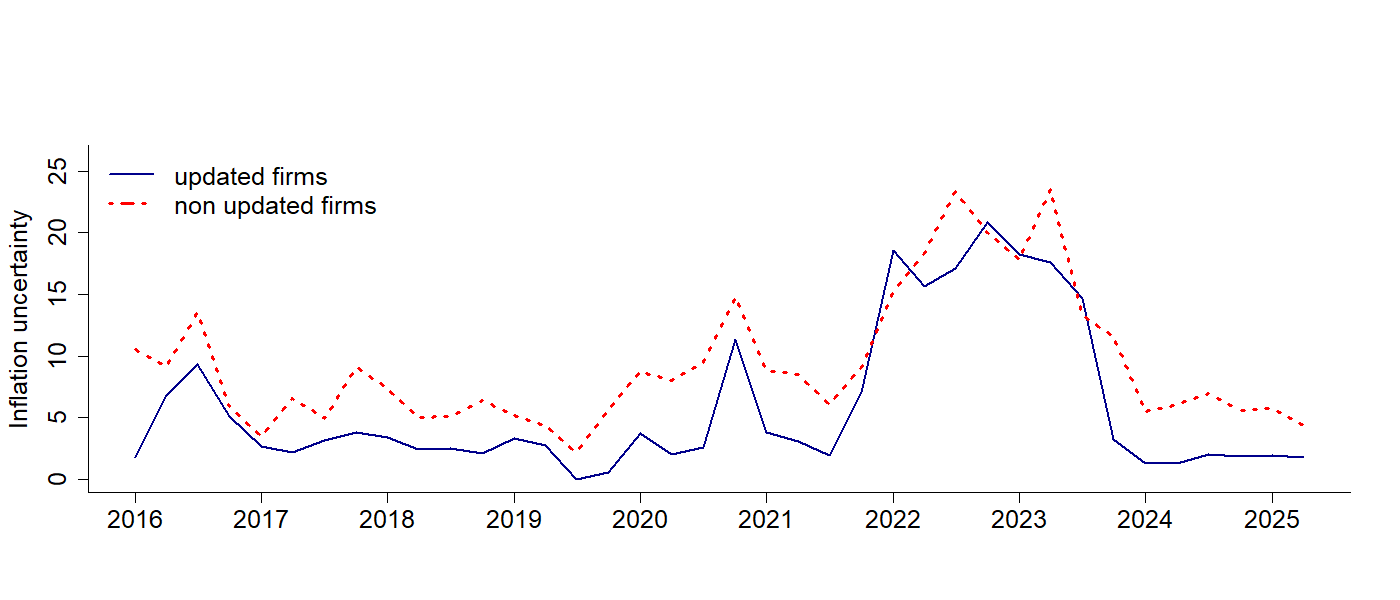}
    \end{subfigure}
       \caption{\scriptsize \justifying \textbf{Inflation Uncertainty Index.} The figure reports the inflation uncertainty index for firms estimated by adopting the framework of \cite{binder2017measuring}. Source: Bank of Italy, Survey on inflation and growth expectations. Sample: 2016:Q1–2025:Q2.}
    \label{figura 2.app}
    \end{minipage}
\end{figure}

\autoref{figura 2.app} plots the inflation-uncertainty index, constructed following \cite{binder2017measuring}, separately for two groups of firms:
(i) those that were informed about last month inflation at the time of the interview;
(ii) those that received no information about inflation;

Uncertainty across subgroups exhibits broadly similar dynamics, and the correlation between the two series is about 88\%. The share of firms classified as uncertain in case (ii) is about 3 percentage points higher on average, but this gap narrows sharply during the post-pandemic recovery. This pattern provides evidence about the fact that the information treatment effectively reduces the level of uncertainty about future inflation. Furthermore, it also suggests that the surge in inflation uncertainty in that period is not primarily driven by informational frictions.

\section{B-PSVAR}\label{App. IVrobust}

\subsection{Daily impulse response functions} \label{app. hf responses}

\begin{figure}[H]
    \centering
    \begin{minipage}{1\linewidth}
        \centering
    \begin{subfigure}[t]{0.32\textwidth}
      \centering
      \includegraphics[width=\linewidth]{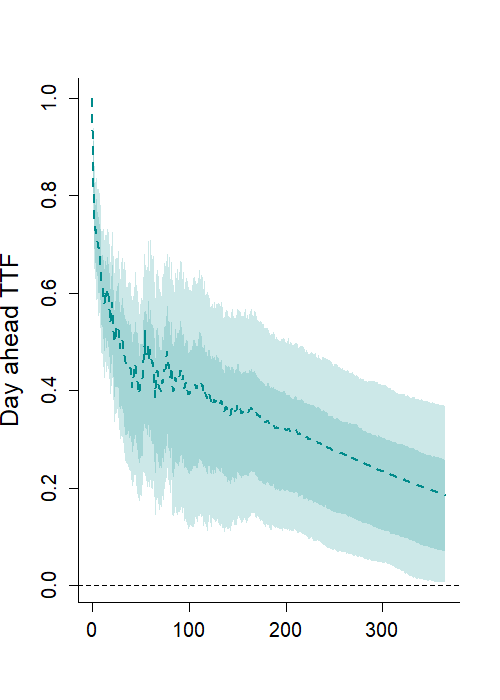}
    \end{subfigure}\hfill
    \begin{subfigure}[t]{0.32\textwidth}
      \centering
      \includegraphics[width=\linewidth]{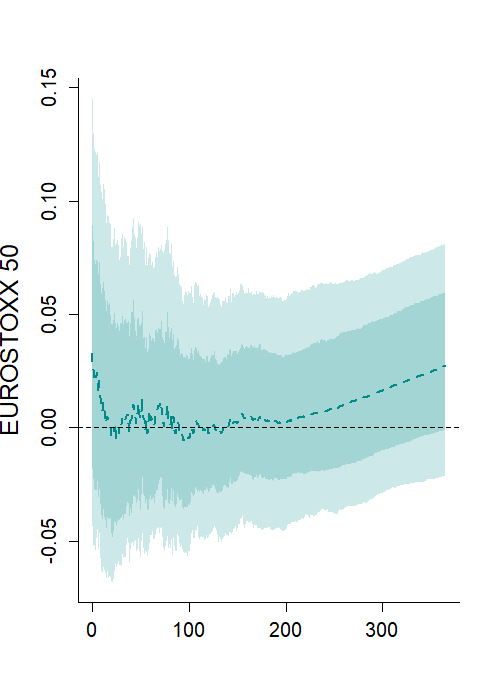}
    \end{subfigure}\hfill
    \begin{subfigure}[t]{0.32\textwidth}
      \centering
      \includegraphics[width=\linewidth]{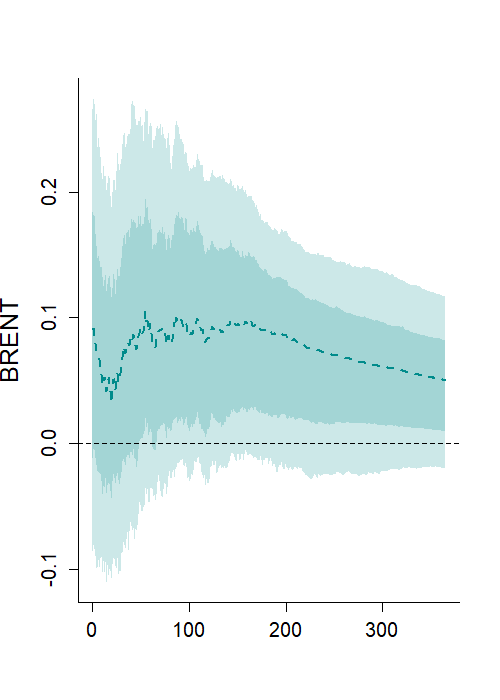}
    \end{subfigure}

    \vspace{0.2em}

    \begin{subfigure}[t]{0.32\textwidth}
      \centering
      \includegraphics[width=\linewidth]{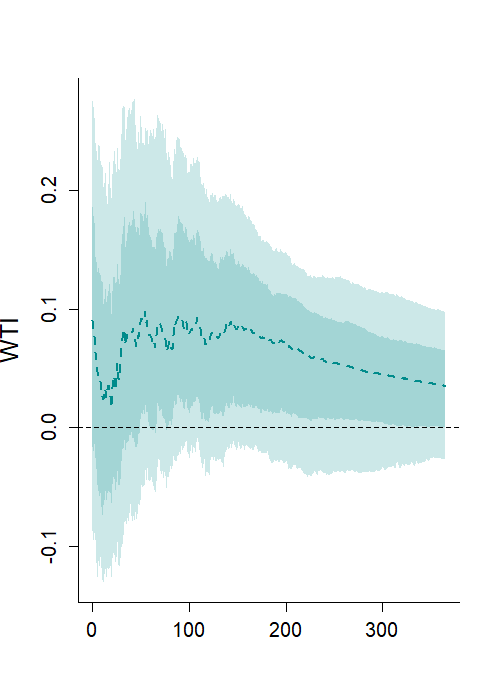}
    \end{subfigure}\hfill
    \begin{subfigure}[t]{0.32\textwidth}
      \centering
      \includegraphics[width=\linewidth]{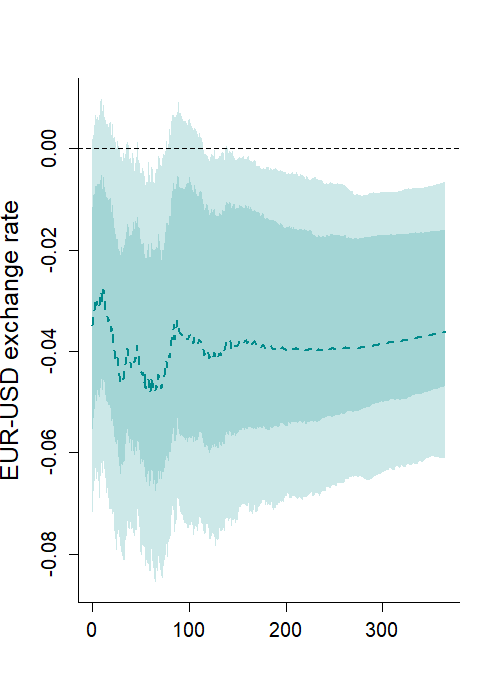}
    \end{subfigure}\hfill
    \begin{subfigure}[t]{0.32\textwidth}
      \centering
      \includegraphics[width=\linewidth]{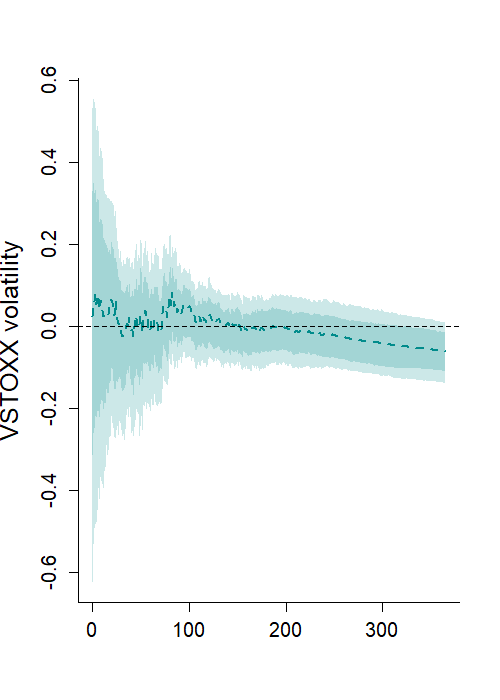}
    \end{subfigure}

    \vspace{0.2em}

    \begin{subfigure}[t]{0.32\textwidth}
      \centering
      \includegraphics[width=\linewidth]{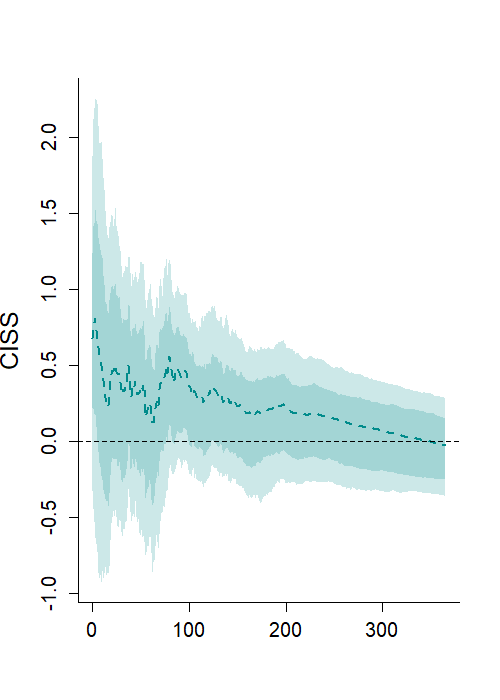}
    \end{subfigure}\hfill
    \begin{subfigure}[t]{0.32\textwidth}
      \centering
      \includegraphics[width=\linewidth]{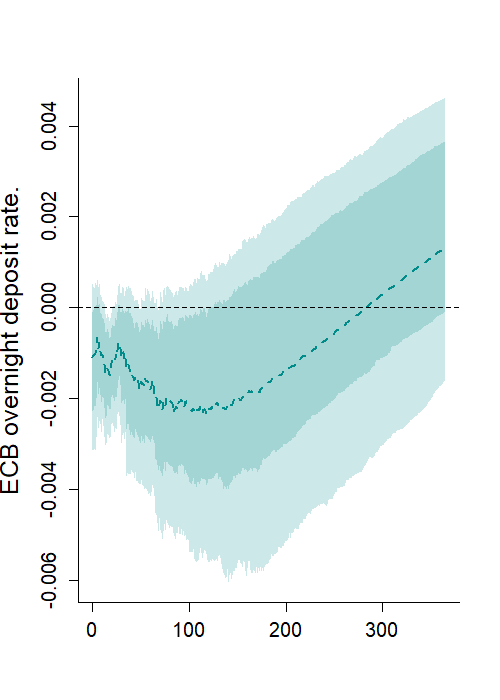}
    \end{subfigure}\hfill
    \begin{subfigure}[t]{0.32\textwidth}
      \centering
      \includegraphics[width=\linewidth]{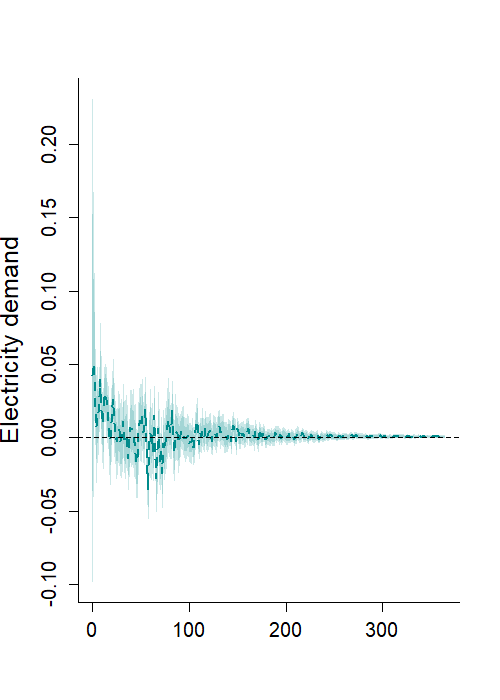}
    \end{subfigure}
    
       \caption{\scriptsize \justifying \textbf{Impulse response functions of energy market variables to a natural gas supply shock.} The figure reports the impulse response functions of the Composite Indicator of Systemic Stress (CISS), the day-ahead TTF natural gas price, the ECB overnight deposit rate, electricity demand, the EURO STOXX 50 index, the EUR-USD exchange rate, the VSTOXX volatility index, and crude oil prices (WTI and Brent) to a natural gas supply shock. Shaded areas represent 68\% and 90\% confidence intervals, computed using a moving block bootstrap procedure. Source: LSEG refinitiv, Copernicus Climate Change Service (2020): Climate and energy indicators for Europe from 1979 to present derived from reanalysis. Sample: 2004:M1–2025:M6.}
    \label{fig:hf responses}
    \end{minipage}
\end{figure}

In \autoref{fig:hf responses} we report the high frequency impulse response functions for our daily VAR-X. The impulse responses display a strong and persistent increase in natural gas prices, while electricity demand does not rise and variables typically associated with global demand show no systematic expansion. 

The muted financial market reactions and the depreciation of the euro are consistent with an adverse natural gas supply shock affecting the European economy. A negative natural gas supply shock worsens the euro area’s terms of trade and growth outlook, leading to a gradual depreciation of the euro through trade balance and capital flow channels, in the absence of an offsetting monetary policy response. The lagged fall in the EUR-USD exchange rate reflect the depreciation of the Euro consequently to the adverse supply shock. The small and statistically insignificant response of electricity demand provides evidence that the identified shock is not driven by demand-related disturbances. 

\subsection{Comparability with Proxy-VAR responses}

We compare our baseline identification strategy based on the bridge Proxy-SVAR with a standard Proxy-VAR approach estimated directly at the monthly frequency. In the latter specification, we use the aggregated version of the high-frequency narrative shock as an external instrument for a monthly VAR, following the conventional proxy identification framework. In this case, the F-statistic is 16.6, while the heteroskedasticity robust F-statistics is equal to 40.

\begin{figure}[H]
    \centering
  
    \begin{minipage}{1\linewidth}
        \centering
    \begin{subfigure}[t]{0.32\textwidth}
      \centering
      \includegraphics[width=\linewidth]{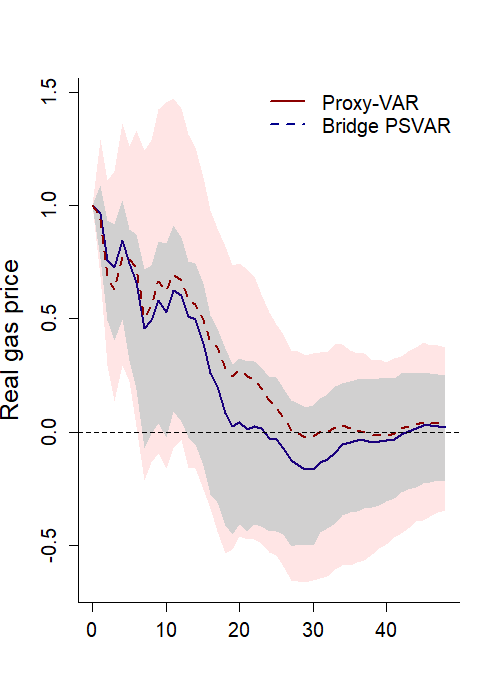}
   
    \end{subfigure}\hfill
    \begin{subfigure}[t]{0.32\textwidth}
      \centering
      \includegraphics[width=\linewidth]{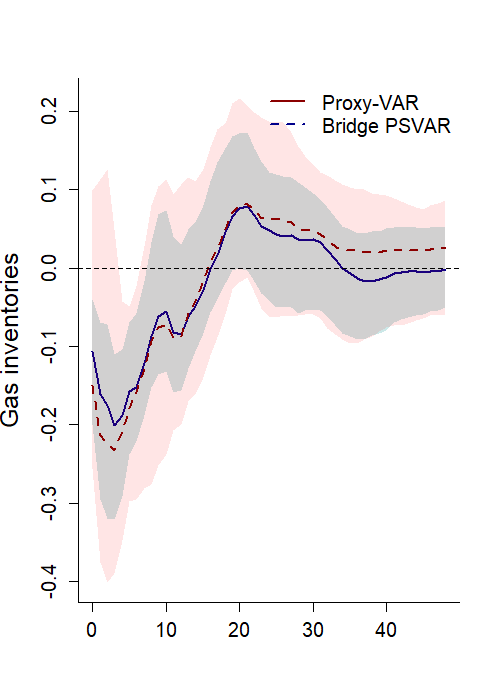}
   
    \end{subfigure}\hfill
    \begin{subfigure}[t]{0.32\textwidth}
      \centering
      \includegraphics[width=\linewidth]{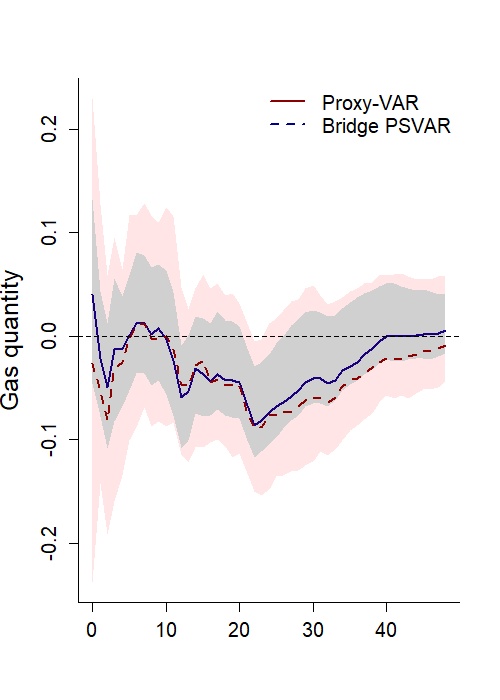}
    
    \end{subfigure}

    \vspace{0.2em}

    \begin{subfigure}[t]{0.32\textwidth}
      \centering
      \includegraphics[width=\linewidth]{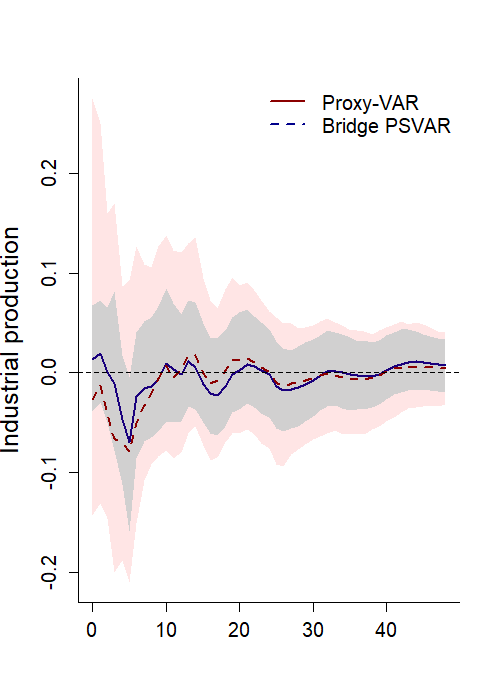}
   
    \end{subfigure}\hfill
    \begin{subfigure}[t]{0.32\textwidth}
      \centering
      \includegraphics[width=\linewidth]{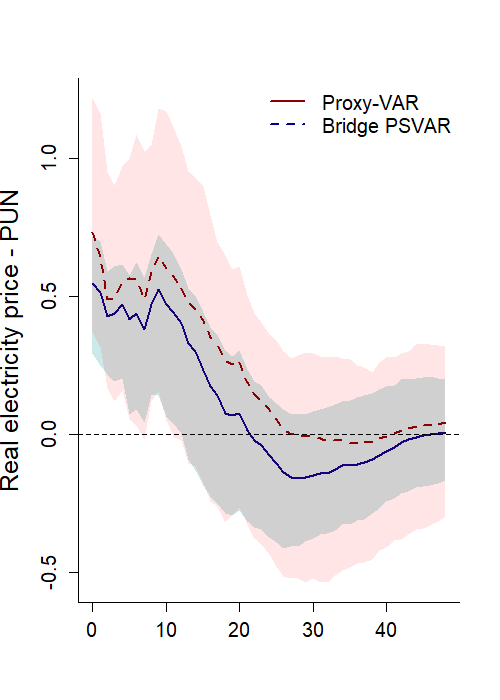}
    
    \end{subfigure}\hfill
    \begin{subfigure}[t]{0.32\textwidth}
      \centering
      \includegraphics[width=\linewidth]{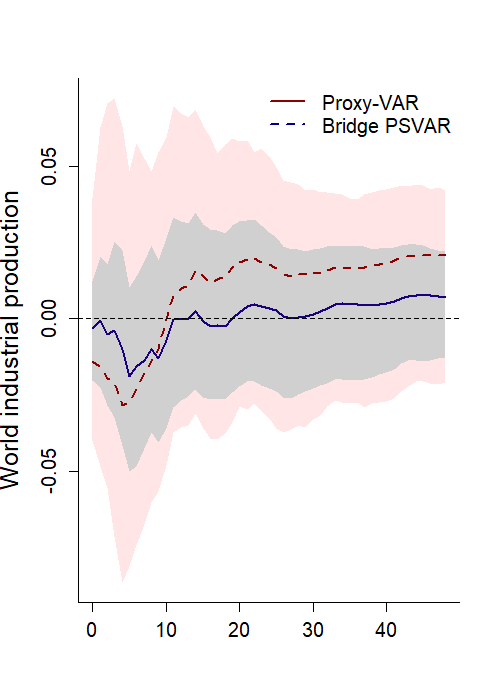}
    
    \end{subfigure}

       \caption{\scriptsize \justifying \textbf{Impulse response functions of energy market variables to a natural gas supply shock.} The figure reports the impulse response functions of natural gas price, gas inventories, gas quantity, industrial production, PUN and world industrial production to a natural gas supply shock. Shaded areas represent 90\% confidence intervals, computed using a moving block bootstrap procedure. Source: Ministero dell’ambiente e della sicurezza energetica, Gestore Mercati Energetici, ISTAT, \cite{BH2019}, \cite{caldara2022measuring}. Sample: 2004:M1–2025:M6.}
    \label{figura comparison}
    \end{minipage}
\end{figure}

\autoref{figura comparison} reports the impulse response functions obtained under the two approaches. Overall, the qualitative responses of energy market variables and real activity are broadly consistent across the two identification strategies. 
However, the responses obtained from the bridge Proxy-SVAR are generally sharper and more precisely estimated, particularly for gas prices, quantities, and downstream variables such as electricity prices and industrial production.

These differences reflect the advantages of identifying the shock at high frequency before aggregation. By exploiting daily information, the bridge approach limits the distortions induced by temporal aggregation and reduces contamination from low-frequency macroeconomic dynamics. As a result, the high-frequency identification yields a cleaner shock, which translates into more informative impulse responses once mapped into the monthly VAR. This comparison highlights the gains from combining high-frequency identification with low-frequency macroeconomic analysis, relative to a standard Proxy-VAR that relies exclusively on aggregated instruments.

\subsection{Controlling for geopolitical risk}

\begin{figure}[H]
    \centering
   
    \begin{minipage}{1\linewidth}
        \centering
    \begin{subfigure}[t]{0.32\textwidth}
      \centering
      \includegraphics[width=\linewidth]{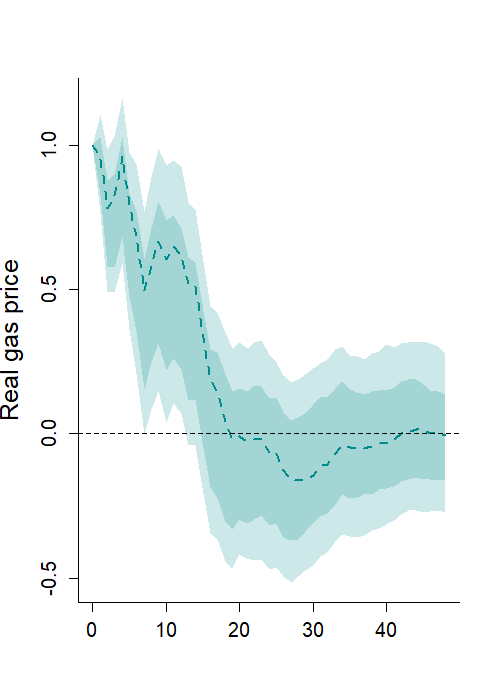}
   
    \end{subfigure}\hfill
    \begin{subfigure}[t]{0.32\textwidth}
      \centering
      \includegraphics[width=\linewidth]{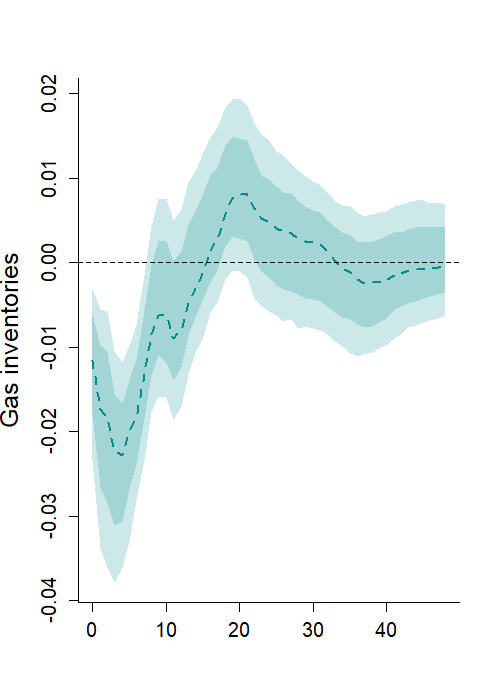}
    
    \end{subfigure}\hfill
    \begin{subfigure}[t]{0.32\textwidth}
      \centering
      \includegraphics[width=\linewidth]{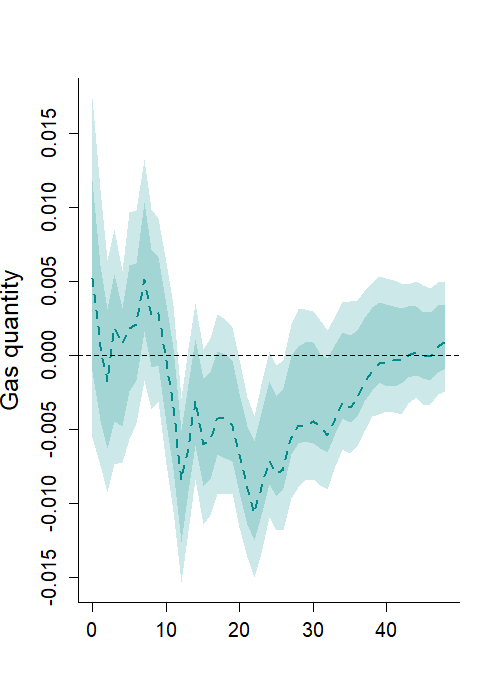}
    
    \end{subfigure}

    \vspace{0.2em}

    \begin{subfigure}[t]{0.32\textwidth}
      \centering
      \includegraphics[width=\linewidth]{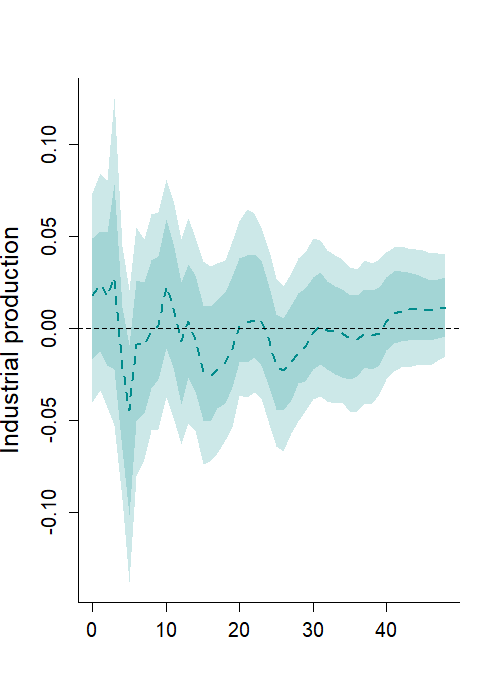}
   
    \end{subfigure}\hfill
    \begin{subfigure}[t]{0.32\textwidth}
      \centering
      \includegraphics[width=\linewidth]{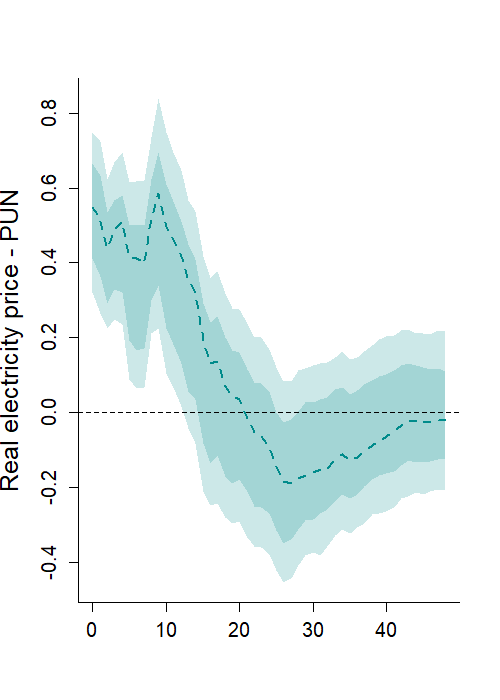}
  
    \end{subfigure}\hfill
    \begin{subfigure}[t]{0.32\textwidth}
      \centering
      \includegraphics[width=\linewidth]{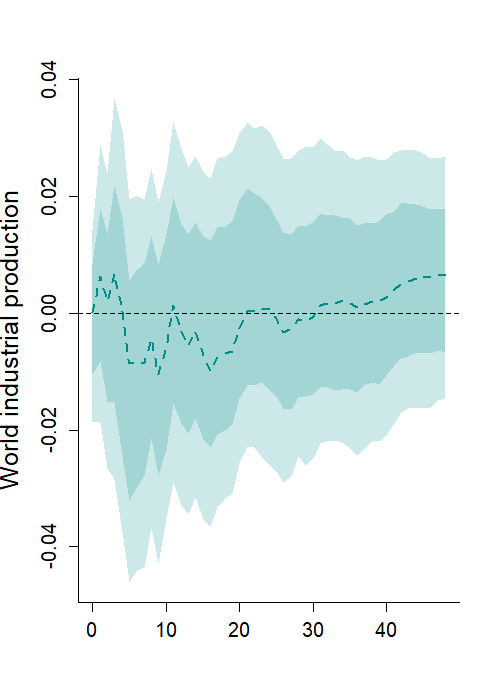}
   
    \end{subfigure}

       \caption{\scriptsize \justifying \textbf{Impulse response functions of energy market variables to a natural gas supply shock.} The figure reports the impulse response functions of natural gas price, gas inventories, gas quantity, industrial production, PUN and world industrial production to a natural gas supply shock. Shaded areas represent 68\% and 90\% confidence intervals, computed using a moving block bootstrap procedure. Source: Ministero dell’ambiente e della sicurezza energetica, Gestore Mercati Energetici, ISTAT, \cite{BH2019}, \cite{caldara2022measuring}. Sample: 2004:M1–2025:M6.}
    \label{figura 11.app}
    \end{minipage}
\end{figure}

\autoref{figura 11.app} shows the impulse response functions obtained by augmenting our energy-sector VAR with the geopolitical risk index of \cite{caldara2022measuring}. The responses are qualitatively unchanged relative to the baseline specification. This result indicates that the instrument is not simply capturing broad geopolitical risk shocks and supports its interpretation as identifying news about disruptions to the physical supply of natural gas.

\subsection{Alternative measure of inflation uncertainty}

\begin{figure}[H]
    \centering
    \begin{minipage}{1\linewidth}
    \centering
    
    \begin{subfigure}[t]{0.32\textwidth}
      \centering
      \includegraphics[width=\linewidth]{plots/Uncertainty_4_gas_shock_IRF_IV.png}
    \end{subfigure}
    \begin{subfigure}[t]{0.32\textwidth}
      \centering
      \includegraphics[width=\linewidth]{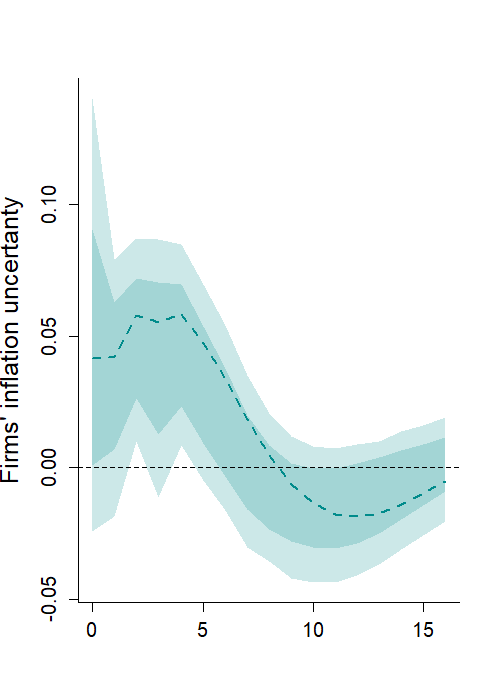}
    \end{subfigure}

    \caption{\scriptsize \justifying \textbf{Impulse response functions of firms' inflation uncertainty to a natural gas supply shock.} The figure reports the impulse response functions of the inflation uncertainty index, constructed by following the two alternative methodologies if \cite{Jurado2015} and \cite{binder2017measuring}, to a natural gas supply shock. Shaded areas represent 68\% and 90\% confidence intervals computed through a moving block bootstrap procedure. Source: Bank of Italy, Survey on inflation and growth expectations. Sample: 2005:Q1–2025:Q2.}
    \label{figura 5 uncertaiinty}
    \end{minipage}
\end{figure}

In \autoref{figura 5 uncertaiinty} we compare the impulse response function of firms' inflation uncertainty, constructed by using the alternative methodologies of \cite{Jurado2015} and \cite{binder2017measuring}, to a a natural gas supply shock. The estimated impulse response functions are similar, providing robustness with respect to the use of alternative proxies of firms' inflation uncertainty. For both of of these measures, the shock produces a lagged hump-shaped response, which starts to revert to zero in about 5 quarters. 

\section{State dependent local projections}\label{App. sdlp}

\subsection{State variable defined on survey rounding measure}

\begin{figure}[H]
    \centering
    \begin{minipage}{1\linewidth}
    \begin{subfigure}[h]{1\textwidth}
    \includegraphics[width=\textwidth]{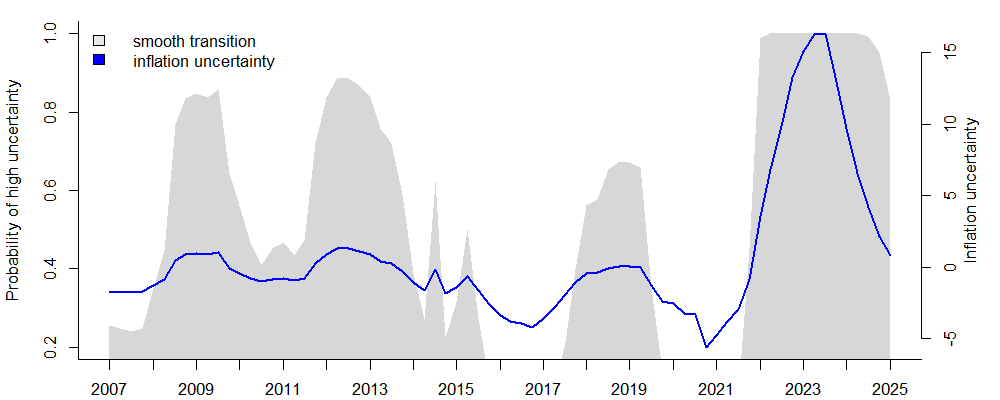}
    \end{subfigure}

    \caption{\scriptsize \justifying \textbf{Inflation uncertainty.} 
    Top panel: estimated probability of high inflation uncertainty following \cite{binder2017measuring}, 
    computed via smooth transition logistic function (shaded areas). 
    Blue line represent the underlying state variable.
    Source: Bank of Italy, Survey on inflation and growth expectations. Sample: 2007:Q1–2025:Q2.}
    \label{figura 6app}
    \end{minipage}
\end{figure}

In this section, we assess the robustness of our results to an alternative measure of firms’ inflation uncertainty. Specifically, instead of relying on the uncertainty index constructed following \cite{Jurado2015}, we use the measure proposed by \cite{binder2017measuring}. We replicate the entire state-dependent local projection analysis using this alternative indicator to define high and low uncertainty regimes. \autoref{figura 6app} reports the estimated smooth transition probabilities.

\begin{figure}[H]
\centering

\parbox{\linewidth}{\centering\footnotesize\textit{Shock coefficient $\tilde{\beta}_h$ by inflation-uncertainty state }}\\
\begin{subfigure}{0.32\linewidth}\centering
    \includegraphics[width=\linewidth,keepaspectratio]{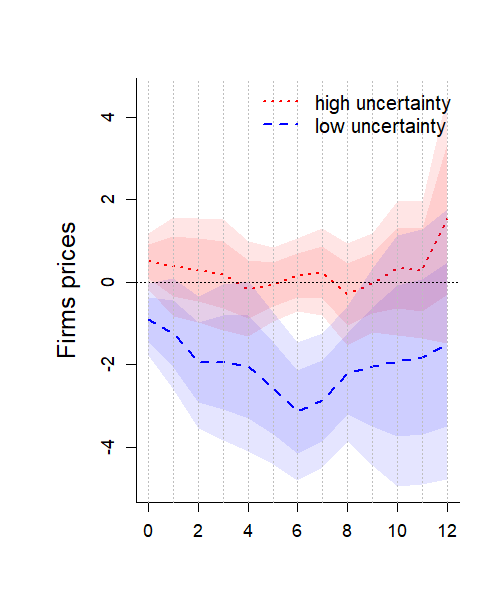}
\end{subfigure}
\begin{subfigure}{0.32\linewidth}\centering
    \includegraphics[width=\linewidth,keepaspectratio]{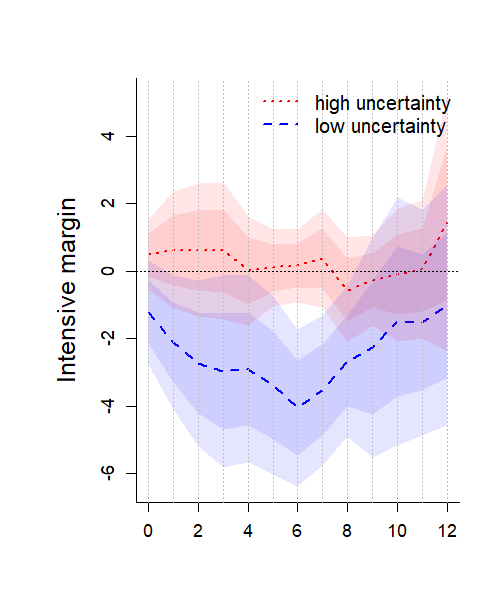}
\end{subfigure}
\begin{subfigure}{0.32\linewidth}\centering
    \includegraphics[width=\linewidth,keepaspectratio]{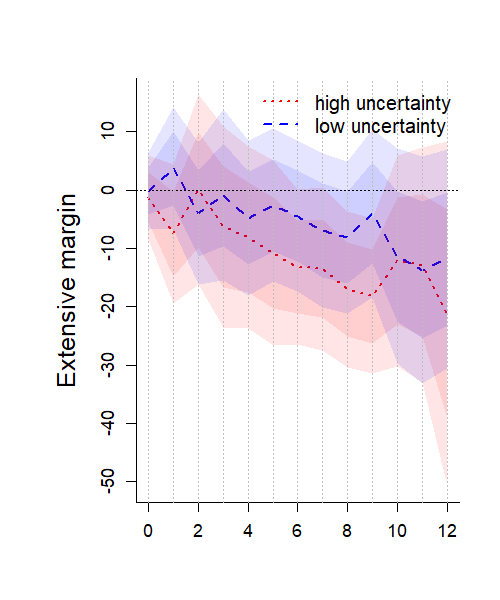}
\end{subfigure}\\[-0.25em]

\parbox{\linewidth}{\centering\footnotesize\textit{Interaction coefficient $\tilde{\gamma}_h$ (shock $\times$ labor market tightness) by inflation-uncertainty state}}\\
\begin{subfigure}{0.32\linewidth}\centering
    \includegraphics[width=\linewidth,keepaspectratio]{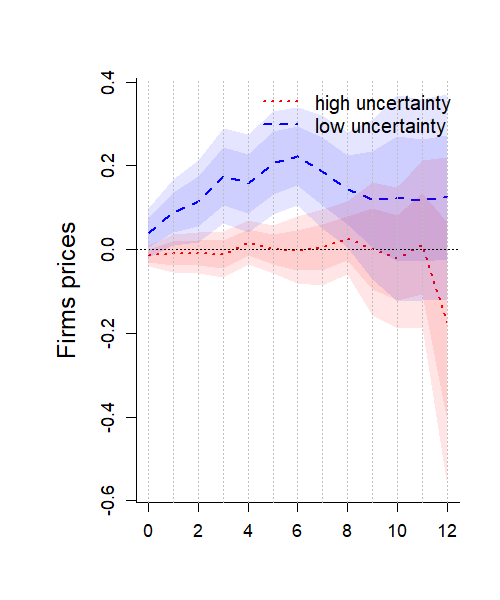}
\end{subfigure}
\begin{subfigure}{0.32\linewidth}\centering
    \includegraphics[width=\linewidth,keepaspectratio]{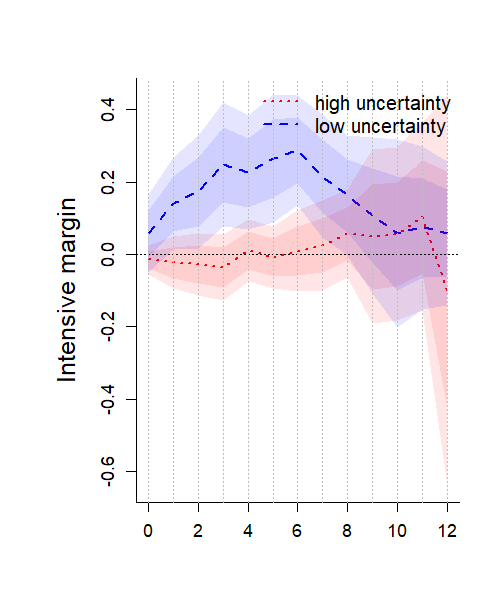}
\end{subfigure}
\begin{subfigure}{0.32\linewidth}\centering
    \includegraphics[width=\linewidth,keepaspectratio]{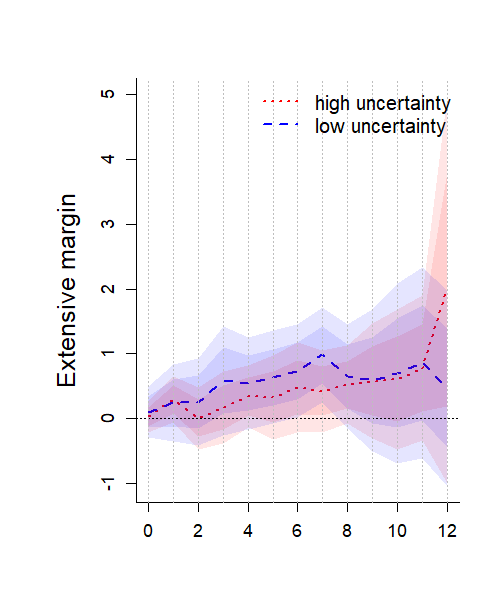}
\end{subfigure}\\[-0.25em]

\parbox{\linewidth}{\centering\footnotesize\textit{Interaction coefficient $\tilde{\delta}_h$ (shock $\times$ ZLB dummy) by inflation-uncertainty state}}\\
\begin{subfigure}{0.32\linewidth}\centering
    \includegraphics[width=\linewidth,keepaspectratio]{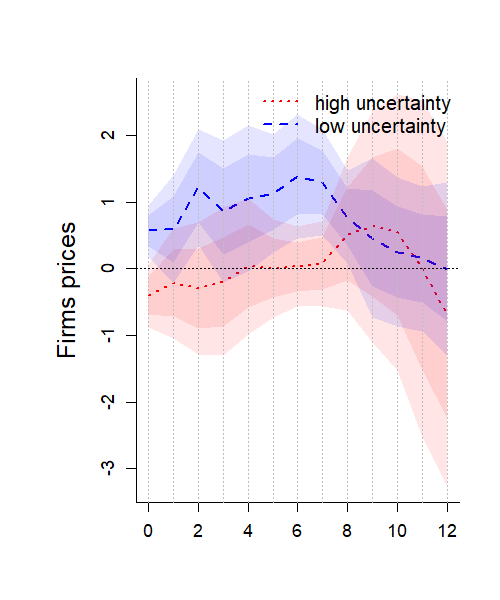}
\end{subfigure}
\begin{subfigure}{0.32\linewidth}\centering
    \includegraphics[width=\linewidth,keepaspectratio]{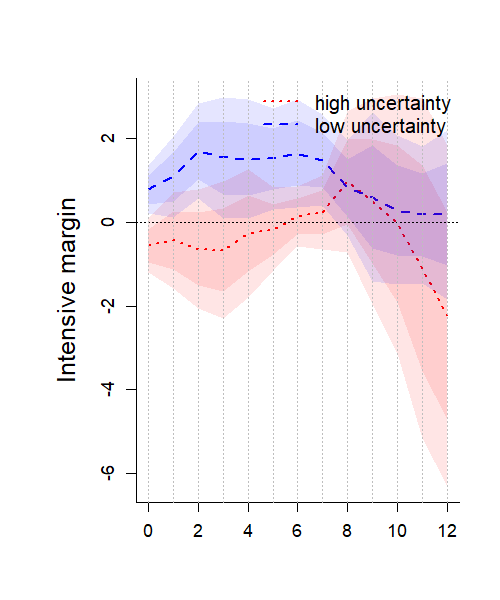}
\end{subfigure}
\begin{subfigure}{0.32\linewidth}\centering
    \includegraphics[width=\linewidth,keepaspectratio]{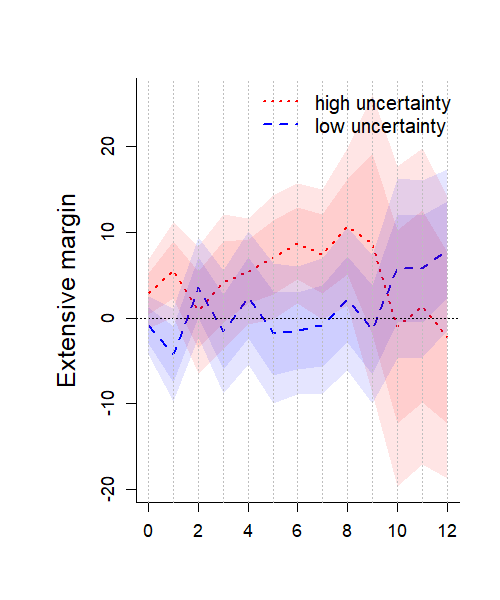}
\end{subfigure}

    \caption{\scriptsize \justifying \textbf{State dependent response of firms' prices and price setting margin.} 
    The figure shows the state-dependent responses of firms’ prices (left column) and the intensive (center column) and extensive (right column) price-setting margins to a natural gas supply shock. The first row reports the estimated coefficient on the shock, $\tilde{\beta}_h$, in high (red) and low (blue) inflation-uncertainty states at each horizon $h$. The second row reports the coefficient on the interaction between the shock and labor market tightness in each uncertainty state. The third row reports the coefficient on the interaction between the shock and the zero lower bound (ZLB) dummy in each uncertainty state. States of high and low inflation uncertainty are built starting from the measure of \cite{binder2017measuring}. Bands denote 68\% and 90\% confidence intervals based on Newey--West standard errors. Source: Bank of Italy, Survey on Inflation and Growth Expectations; ISTAT, EUROSTAT.\emph{Sample: 2007Q1–2025Q2.}}

    \label{figura 7binder}
\end{figure}

The resulting impulse response functions are reported in \Cref{figura 7binder}. The responses are virtually indistinguishable from those obtained under the baseline specification, both in terms of magnitude and statistical significance. In particular, the stronger price response under low uncertainty, the dominant role of the intensive margin, and the state-dependent amplification associated with labor market tightness and the zero lower bound are all preserved.

\end{document}